\documentclass[aps,pre,floatfix,superscriptaddress,reprint]{revtex4-2}

\usepackage{graphicx,psfrag,color}
\usepackage{bm,bbm}

\usepackage{amsmath,amssymb}

\usepackage{comment}
\excludecomment{mycomments}

\newcommand{\average}[1]{\left<{#1}\right>}

\newcommand{\bp}{\bm{p}}
\newcommand{\bW}{\bm{W}}

\newcommand{\E}{\mathrm{e}}
\newcommand{\D}{\mathrm{d}}

\newcommand{\up}{ \uparrow}
\newcommand{\down}{\downarrow}

\newcommand{\Tp}{T_+}
\newcommand{\Tm}{T_-}
\newcommand{\peq}{p^{eq}}
\newcommand{\bpeq}{\bm{p}^{eq}}
\newcommand{\Epm}{\average{E}^{eq}_+}
\newcommand{\Emm}{\average{E}^{eq}_-}
\newcommand{\Esqpm}{\average{E^2}^{eq}_+}
\newcommand{\betafin}{\beta_{f}}
\newcommand{\betain}{\beta_{i}}
\newcommand{\Tfin}{T_{f}}
\newcommand{\Tin}{T_{i}}

\newcommand{\p}[1]{\left({#1}\right)}
\newcommand{\pq}[1]{\left[{#1}\right]}
\newcommand{\pg}[1]{\left\{{#1}\right\}}

\newcommand{\ders}[1]{\frac{d^2 }{d #1^2}}

\newcommand{\alb}[1]{{ \color{blue} #1}}
\begin{document}
\title{On thermalization of a system with discrete phase space}

\author{Alberto Imparato}
\address{Department of Physics, University of Trieste, Strada Costiera 11, 34151 Trieste, Italy}
\address{ Istituto Nazionale di Fisica Nucleare, Trieste Section, Via Valerio 2, 34127 Trieste, Italy}

\date{\today}
\begin{abstract}
We investigate the thermalization of a stochastic system with discrete phase space, initially at equilibrium at temperature $T_i$ and then termalizing in an environment at temperature $T_f$, considering both cases  $T_i>T_f$ and $T_i<T_f$. For the simple case of a system with constant energy gaps, we show that the relation between the time scales of the cooling and heating processes is not univocal, and depends on the magnitude of the energy gap itself. Specifically the eigenvalues of the corresponding stochastic matrix set the time scales of the relaxation process and for large energy gaps the cooling process is found to exhibit the shortest relaxation times to equilibrium while the heating process is found to be faster at all scales for small gaps.
We consider both the Kullback--Leibler divergence and the Fisher information and its related quantities to quantify the degree of thermalization of the system. In the intermediate to long time regime both quantities are found to bear the same type of information concerning the rate of thermalization, and follow the ordering predicted by the dynamic eigenvalues.
We then consider a more complex system with a more intricate stochastic matrix, namely a 1D Ising model,  and confirm the findings on the existence of two regimes, one in which cooling becomes faster than heating.

We make contact with a previous work where an harmonic oscillator was used as working fluid and the heating process was always found to be faster than the cooling one.
\end{abstract}
\maketitle

Thermalization refers to the process by which a physical system evolves towards thermal equilibrium when in contact with a surrounding environment at constant temperature. 
It arises across diverse contexts, from quantum systems to biological processes and remains a subject of active research due to its fundamental and often unexpected features.
A typical example is the Mpemba effect \cite{mpemba_original}, that refers to the counterintuitive phenomenon where, under certain conditions, a system initially prepared at a higher temperature equilibrates faster than an otherwise identical system starting at a lower temperature. This effect challenges the intuitive expectation that cooler systems should reach equilibrium more quickly and has been observed in both classical and quantum systems \cite{klich_mpemba,lucon2022mpemba,holtzman2022mpemba,das_mpemba}.

In stochastic systems initially prepared  in a non equilibrium state and then coupled to a bath at {\it final} temperature $T_f$ the thermalization process occurs along different pathways that are selected by the initial conditions, and it involves energy exchange with the bath and transient currents across the system at different length and time scales. Thus, while the approach to equilibrium may appear straightforward, and governed by few intuitive mechanisms, it often reveals rich dynamical behavior and abundance of transient phenomena.

An intriguing problem to investigate is the timescale over which thermalization occurs and the key features that drive this process. To explore this, one requires a set of operational tools capable of quantifying both the extent of equilibration and the associated timescales.

In this spirit, in \cite{Ito2020}, the authors introduced the concept of statistical velocity and statistical length for a stochastic process to study the constraints on the rates of observables of interest.
 As a result the authors found a speed limit on the evolution of
stochastic observables involving the statistical velocity defined as the square root of the Fisher information.
Introduced in the context of information geometry, Fisher information on turn quantifies the rate of change of the Kullback--Leibler (KL)  divergence along a stochastic trajectory.
Recently  the information geometry approach has been extended to the to quantum stochastic thermodynamics,  in particular in \cite{Bettmann2025} the concept of Fisher information has been extended to the quantum realm. 

In \cite{Ibanez2024} the authors  used the definition of statistical length (or distance) introduced in \cite{Ito2020} for stochastic dynamics to argue that in two opposite thermalization processes between two temperatures $\Tp$ and $\Tm$, with $\Tp> \Tm$, the heating process is always faster than the reverse cooling process. 
In this framework, the statistical length of a process is simply the integral of the instantaneous statistical velocity, thus suggesting the 
wording  "thermal kinematics'', for its resemblance with the kinematics in classical mechanics.
Specifically the authors of \cite{Ibanez2024} use a classical harmonic oscillator as working fluid for their theoretical and experimental  
study of the thermalization processes between two temperatures $\Tp$ and $\Tm$. In the heating process the system is initially at equilibrium at temperature $\Tm$ for $t\le 0$, and at $t>0$ is put in contact with a bath at temperature $\Tp$. In the reverse (cooling) process the system is initially at equilibrium at temperature $\Tp$  and at $t>0$ is put in contact with a bath at temperature $\Tm$. The statistical lengths introduced in \cite{Ito2020} turns out to be symmetric, i.e. equal for the two processes, with the heating process covering this {\it distance} in a shorter time thus for the harmonic oscillator heating turns out to be faster than cooling, as one would intuitively expect.
It is worth noting that an analogous result has been obtained in the quantum case in \cite{Tejero2025} where a two-level system, the quantum harmonic
oscillator, and a trapped quantum Brownian particle showed faster heating than cooling under appropriate conditions.

In the present paper, we take a different approach, and consider classical systems with discrete phase space that evolve through a master equation and undergo thermalization processes between two temperatures  $\Tp>\Tm$.
This allows us to distinguish between the different time scales corresponding to the eigenvalues of the stochastic matrix setting the transition rates between the different states. Such time scales must enter into any possible kinematic description of an equilibration process. 

The  main result of the present paper is that for a given system with discrete phase space, one can always find a regime where the heating process become slower than the cooling. This occurs in a regime where the temperatures are low or energy gaps between the states are large compared to the thermal energies $\Tp$ and $\Tm$ set by the two baths (here and in the following we set $k_B=1$).
Furthermore, the statistical length introduced in \cite{Ito2020,Ibanez2024} turns out to be  asymmetric between the two processes, and in general the heating is characterized by a longer statistical length, thus signaling different thermalization pathways between the heating and the cooling processes. The symmetry of the statistical length is restored for two paradigmatic systems: a  two state system (e.g. a single spin) and the classical harmonic oscillator, the latter in accordance with \cite{Ibanez2024}.

In terms of the different stochastic matrix eigenvalues describing the convergence towards equilibrium it is know that the first eigenvalue is vanishing, corresponding to the equilibrium state, while the second slowest eigenvalue describes the global decay towards equilibrium, all the other eigenvalues describing faster processes occurring in the system during the relaxation towards equilibrium \cite{vanKampen1981}.
The main result of the present paper can be thus rephrased as follows: in the limit where the dynamics becomes slow, i.e. for low temperatures or large energy gaps, the relaxation rates towards equilibrium, given by the second and higher order slowest eigenvalues, can become smaller for a heating process than for the corresponding cooling process,  resulting in a somehow counterintuitive effect, as one would expect the dynamics driven by the hot reservoir being faster than the one driven by the cold reservoir.
\begin{mycomments}
\alb{perché avevo scritto low temperatures? perché ho scritto che diventa slow?}
\end{mycomments}

\section{Stochastic system}
\label{sec:I}
\begin{mycomments}
\alb{vedi sistemi eccitonici di plenio et al}
\end{mycomments}

In the following we use the standard formalism of the master equation, and focus on relaxation processes between an initial ($t\le 0$) and a   final ($t>0$) temperature, $\Tin$ and $\Tfin$, respectively. The theory of the master equation is  briefly reviewed, and we refer the reader to, e.g., \cite{vanKampen1981}, for a more detailed discussion.

We consider a system with $N$ discrete energy levels $\epsilon_k$, and we order the energy levels in ascending order $\epsilon_0=0\le \epsilon_1\le \dots \le \epsilon_{N-1}$. 

\begin{figure}[h]
\center
\psfrag{ }[ct][ct][1.]{ }
\includegraphics[width=8cm]{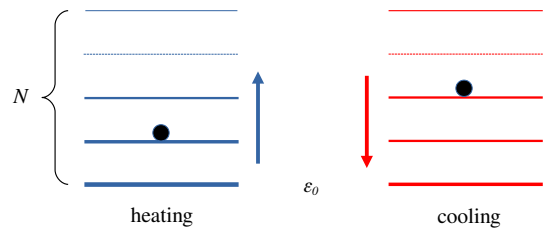}
\caption{Initial state for the heating (left) and the cooling (right) process  for a system with $N$ discrete energy levels. The different thickness of the single levels represents the different size of the initial populations in each level (not in scale). For $T_i=T_-$ the lower energy levels are more populated with respect to the case $T_i=T_+$. The arrows represent the average directions of the initial probability currents in the two cases.  }
\label{sistema:fig}
\end{figure}

%Following \cite{Ibanez2024},
In this paper we want to study the thermalization dynamics between two temperatures $\Tm$ and $\Tp$. More precisely, we compare the relaxation from $\Tm$ to $\Tp$, where the system, initially at equilibrium at $\Tin=\Tm$ ($t\le0$)  is placed in contact with a bath at $\Tfin=\Tp$ ($t>0$), this corresponds to the heating protocol. In the reverse protocol the system is initially at equilibrium at temperature $\Tin=\Tp$ ($t\le0$) and then cooled by placing it in contact with a bath at temperature $\Tfin=\Tm$ ($t>0$).
The initial states of the two processes are depicted in fig.~\ref{sistema:fig}.

We ask the following question: in which process is the new equilibrium thermal state reached first?

In the following we will use the notation $\bp_-^+(t)$ and  $\bp_+^-(t)$ that indicate the probability vector for the heating and cooling process, respectively (from initial to final temperature $\bp_i^f(t)$).

The dynamics is described by a master equation of the type
\begin{equation}
\dot \bp(t)=\bW \cdot \bp(t),
\label{eq:mast}
\end{equation} 
where $\bW$ is a stochastic matrix with elements $w_{ij}$, whose dependence on the (inverse) temperature is encoded through the detailed balance condition for its elements
\begin{equation}
\label{det:bal}
w_{kj}=w_{jk} \E^{\beta (\epsilon_j-\epsilon_k)}. 
\end{equation} 
Thus we indicate with $\bW^+$ and $\bW^-$ the stochastic matrices ruling the dynamics during the heating and the cooling process, respectively.

Let $\lambda_n$, with $n=0,\dots, N-1$ be the eigenvalues of $\bW$ for a given temperature, and let's assume they have been ordered in ascending order $\lambda_0>\lambda_1\ge \dots \lambda_{N-1}$. 
\begin{mycomments}
\alb{Teza in \cite{Teza2023} claims that the eigenvalues of a matrix with detailed balance are always reals, and this is discussed in the paper \cite{Raz2016}. }
\end{mycomments}
Given that $\bW$ is a stochastic matrix,  the largest eigenvalue is $\lambda_0=0$, with all the other eigenvalues which are negative, see Ch.V in \cite{vanKampen1981}. Thus if $\mathbf{\Phi}_0$ is the eigenvector corresponding to the eigenvalue $\lambda_0$ then it corresponds to the equilibrium thermal probability at the final inverse temperature,  $\mathbf{\Phi}_0 =\bpeq(\betafin)$, while the other eigenvalues $\lambda_n, \, n>0$ determine the rate of relaxation towards the equilibrium state $\bpeq(\betafin)$.

Thus we have 
\begin{equation}
\bp_-^+(t)=\E^{t\bW^+ }  \cdot\bpeq_-\, , \qquad   \bp_+^-(t)=\E^{t \bW^-} \cdot \bpeq_+,
\end{equation} 
with $\bpeq_{\pm}=\bpeq(\beta_\pm)$.

To analyse the difference in the dynamics of the two processes, the heating and the cooling one, one has at least two options.

i) 
One is  tempted to use the KL divergence (or relative entropy) 
\begin{equation}
D(\bp_a|| \bp_b)=\sum_k p_{a,k} \log\p{\frac{p_{a,k}}{p_{b,k}}},
\label{KL:def}
\end{equation} 
to evaluate the convergence toward equilibrium, and evaluate $D(\bp_i^f(t)|| \bpeq_f)$ at any time $t\ge 0$.
However, as detailed in \cite{Ito2020, Ibanez2024}, the KL divergence is not a metric in the space of the probability vectors: it does not, for example, satisfy the triangle inequality, and thus cannot be used as a distance between probabilities, in particular it does not provide a proper {\it distance} from equilibrium.

\begin{mycomments}
\alb{Taken from \cite{Ibanez2024}, see what \cite{Ito2020} says on this}
\end{mycomments}

Nevertheless, the KL relative entropy can provide useful information on the convergence rate of the equilibration process, as we will see in the following.
Let us define the KL distances for the two processes
\begin{equation}
D_-^+(t)=D( \bp_-^+(t)|| \bpeq_+)\, , \qquad D_+^-(t)=D( \bp_+^-(t)|| \bpeq_-).
\label{KL:def1}
\end{equation} 
In appendix \ref{appendix:one} we prove the following inequality at initial time $ D_+^-(0)\ge D_-^+(0)$.
We also prove that, as one would expect, the convergence to zero of the KL divergence  is dominated by the second slowest eigenvalue $\lambda_1$: 
\begin{equation}
D(\bp_i^f(t)|| \bpeq_f) \underset{t\to \infty} {\sim}\E^{2 \lambda_1 t}.
\label{D:longt}
\end{equation} 
Thus the order relation between $D_-^+(t)$ and $ D_+^-$ at long time is dictated by the relation between  $\lambda_1^+$ and   $\lambda_1^-$ which as we will see is not univocally fixed.

ii) One can  use the approach introduced in \cite{Ibanez2024} and consider the Fisher information, the instantaneous statistical velocity and the statistical length that read
\begin{eqnarray}
I(t)&=&\average{ (\partial_t \ln \bp(t))^2},\label{IT:def}\\
v(t)&=&\sqrt{I(t)},\label{vT:def}\\
\mathcal{L}(t)&=&\int_0^t\, \D \tau v(\tau), \label{LT:def}
\end{eqnarray}  
respectively.

The Fisher information  is related to the KL divergence through the relation 
\begin{eqnarray}
D(\bp(t+ \delta t)|| \bp(t))&=&\delta t^2 \sum_k \frac{(\partial_t p_k(t))^2}{p_k(t)} + O(\delta t^3)\nonumber \\
&=&\delta t^2 I(t) + O(\delta t^3)
\label{DL:I}
\end{eqnarray} 
Therefore inspection of eqs.~\eqref{vT:def} suggests that the instantaneous statistical velocity  is related to the rate of change of the KL  divergence.
We notice that in \cite{Ibanez2024}  the long time limit of the  statistical length was found to be  independent of the thermodynamic {\it direction}, i.e., $\mathcal{L}^+_-(\infty)=\mathcal{L}_+^-(\infty)$.
For the systems with discrete energy spectrum considered in the present manuscript, we anticipate that in general this is the case only for $N=2$ states, or in the limit of small energy gaps.
% and $N\to \infty$ thus reconciling our findings with those of  \cite{Ibanez2024} for the classical harmonic oscillator in a heath bath. 

\begin{mycomments}
\alb{Do \cite{Ibanez2024} and \cite{Ito2020} state explicitly that $\mathcal{L}$ is a metric?
Because our result on $\mathcal{L}^+_-(\infty)\neq \mathcal{L}_+^-(\infty)$ contradict that.
Yes, \cite{Ibanez2024} says that before eq. 3}
\end{mycomments}

In appendix \ref{appendix:one} we also prove that the long time behavior of the  Fisher entropy  ( the statistical velocity ) is dominated by the largest eigenvalue of the stochastic matrix $I(t)\sim \E^{2 \lambda_1 t}$ ($v(t)\sim \E^{\lambda_1 t}$). Thus, as for the case of the KL divergence, we will see that the order relation between  $I_-^+(t)$ and $ I_+^-(t)$, and thus between $v_-^+(t)$ and $v_+^-(t)$ is not univocally fixed.
Furthermore, given that the long time behaviour of the KL divergence and the  Fisher entropy are both governed by  $\lambda_1$, there is in principle no reason to prefer one quantity over another to describe the convergence to equilibrium.

We also find that, for all the model considered in this paper, and for a wide set of parameters, the following inequality folds $I_-^+(0)\ge I_+^-(0)$. While the author of this paper has not succeeded in finding a general analytic proof of this inequality besides the case $N=2$, a fair level of intuition can be gained by considering that the following equalities holds

\begin{eqnarray}
&& \left. \frac{d}{d t} D(\bp_+^-(t) || \bpeq_+)\right|_{t=0} =\left. \frac{d}{d t} D(\bp_-^+(t) || \bpeq_-)\right|_{t=0}=0,\nonumber \\
&& I_-^+(0)= \left. \ders{t} D(\bp_+^-(t) || \bp_+^-(0))\right|_{t=0}, \nonumber  \\
&& I_+^-(0)=\left. \ders{t} D(\bp_-^+(t) || \bp_-^+(0))\right|_{t=0}. \nonumber
\end{eqnarray} 
where we notice that $\bp_+^-(0)=\bpeq_+$ and $\bp_-^+(0)= \bpeq_-$.
Thus the inequality $I_-^+(0)\ge I_+^-(0)$ expresses the intuitive expectation that the system departs from its initial condition faster when in contact with the hot bath at temperature $T_+$.

A few comments on the transition rates in the stochastic matrices are now in order.
As far as the symmetric part of the matrix elements $w_{kj}$ is concerned, in stochastic thermodynamics it is customary to consider those terms as temperature independent $w_{kj}=\omega_0 f(k,j)$, with the asymmetric term $f(k,j)$ satisfying the detailed balance condition \eqref{det:bal}. 
We will term the bath corresponding to this choice standard bath (SB).
However, when one derives the dynamical equations starting from a detailed bath-system model, the {\it pre-factor} turns out to depend explicitly on the temperature.
If one starts from a microscopic model of bath interacting with the system of interest, as one standardly does when studying open quantum systems, the transition rates exhibit prefactors that depend explicitly on the temperature, and for a bosonic bath the transition rate between a state $j$ and a state $k$ reads \cite{Breuer02}
\begin{equation}
\omega_{kj}=\frac{\omega_0 |\Delta E_{k j}|}{1-\E^{-\beta |\Delta E_{k j}|}} \begin{cases}
			 \E^{-\beta \Delta E_{k j}}, & \text{if $\Delta E_{k j}>0$ }\\
            1, & \text{otherwise}
		 \end{cases}
\label{omega:Bos}
\end{equation} 
with the symmetric prefactor that goes to $\omega_0/\beta$ in the classical limit $\beta\to 0$ \cite{Zwanzig2001}.
We will term the bath corresponding to this choice bosonic bath (BB), and in the following we will set the time scale by taking $\omega_0=1$ for both types of baths.

\section{Equispaced energy levels}

In the following we consider the case of equispaced energy levels $\epsilon_k=k \epsilon$, with $\epsilon>0$ and $k=0, \dots, N-1$. We also use the notation $\lambda_{N,k}$ to enumerate the eigenvalues of the corresponding $N\times N$ stochastic matrix $\bW$.

We will consider stochastic  block tridiagonal matrices, i.e. matrices whose non-zero elements are located on the lower diagonal, the main diagonal and the upper diagonal, with $w_{kj}=0$ if $|k-j|>1$. This is the case, for example,  of the master equations for a harmonic oscillator where the particle interacts with the bath through the $x$ coordinate ($a$ and $a^\dagger$ operators) or of a single spin system where the interaction  with the bath occurs through the $x$ (or $y$) component of the spin ($\sigma^+$ or $\sigma^-$ operators). 
\begin{mycomments}
\alb{true only for a single spin }
\end{mycomments}

The stochastic matrix thus reads
\begin{equation}
\bW=
\begin{pmatrix}
 -w^\up&  w^\down & 0& 0& \dots& 0 \\
 w^\up &  -(w^\down+w^\up) &  w^\down  & 0& \dots & 0  \\
 0&   w^\up & -(w^\down+w^\up) &  w^\down & \dots & 0  \\
 0&    \vdots  &  \vdots &  \vdots& \ddots & 0  \\
 0&  \dots  & \dots &  0 & w^\up & -w^\down   \\
\end{pmatrix}
\label{W:eq}
\end{equation}

The case of system with degenerate energy levels and transition rates  $w_{kj}$ 
 connecting two arbitrary  microscopic states $k$ and $j$ will be considered in section \ref{deg:sec} .

For the SB we will set $w^\down=\omega_0$ and $w^\up=\omega_0 \exp(-\beta \epsilon)$, while for the BB eq.~\eqref{omega:Bos} gives
\begin{eqnarray}
w^{\up}&=&=\frac{\omega_0 \epsilon}{1-\E^{-\beta\epsilon}} \E^{-\beta\epsilon}\\
w^{\down}&=&=\frac{\omega_0 \epsilon}{1-\E^{-\beta\epsilon}}
\end{eqnarray} 

\begin{figure*}[t]
\center
\psfrag{ }[ct][ct][1.]{ }
\includegraphics[width=0.49\textwidth]{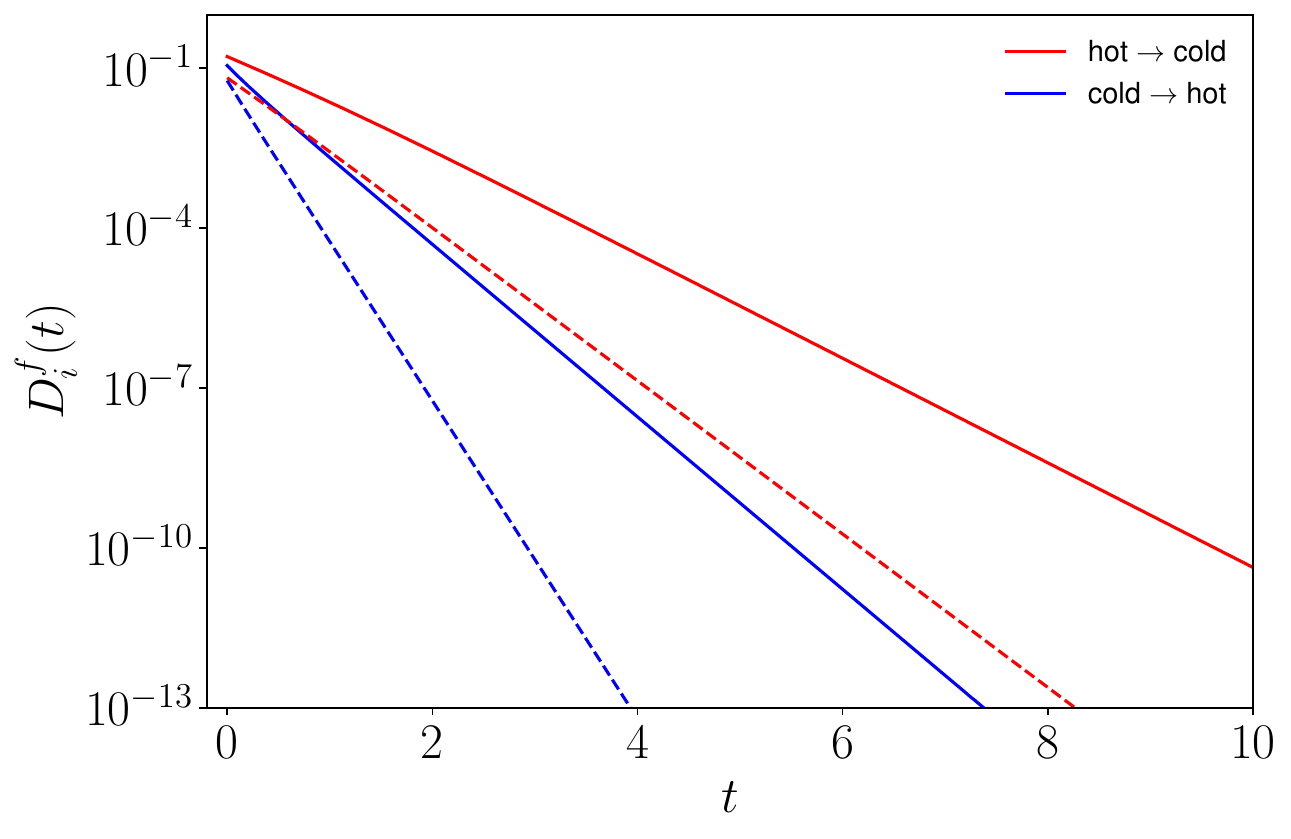}
\includegraphics[width=0.49\textwidth]{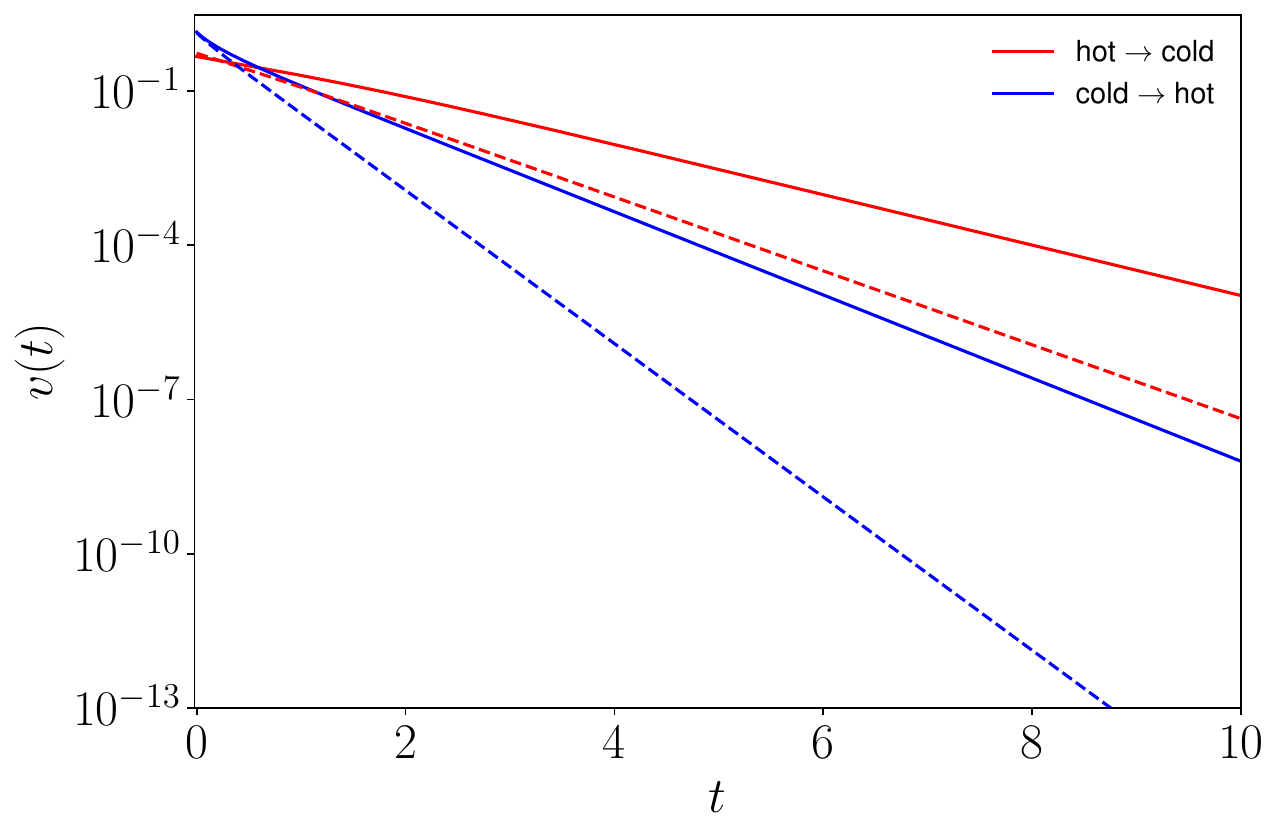}
\includegraphics[width=0.49\textwidth]{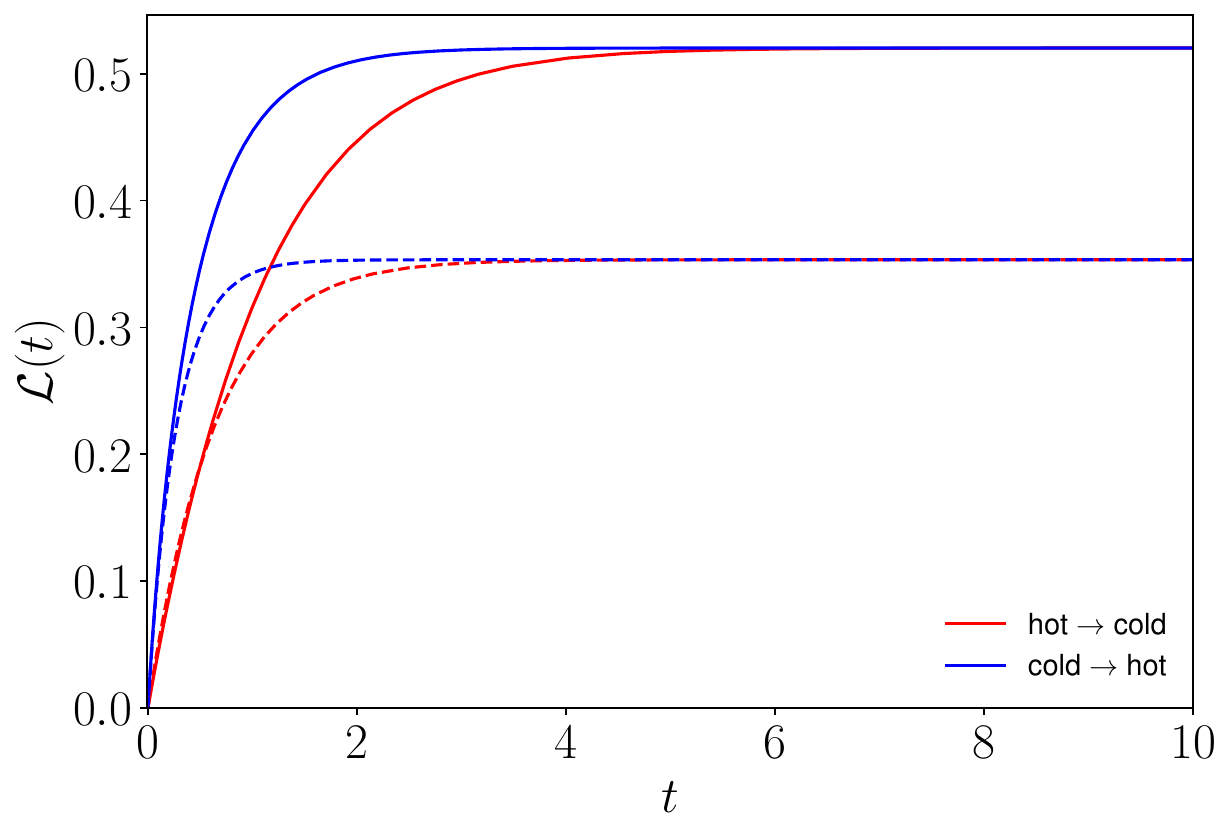}
\includegraphics[width=0.49\textwidth]{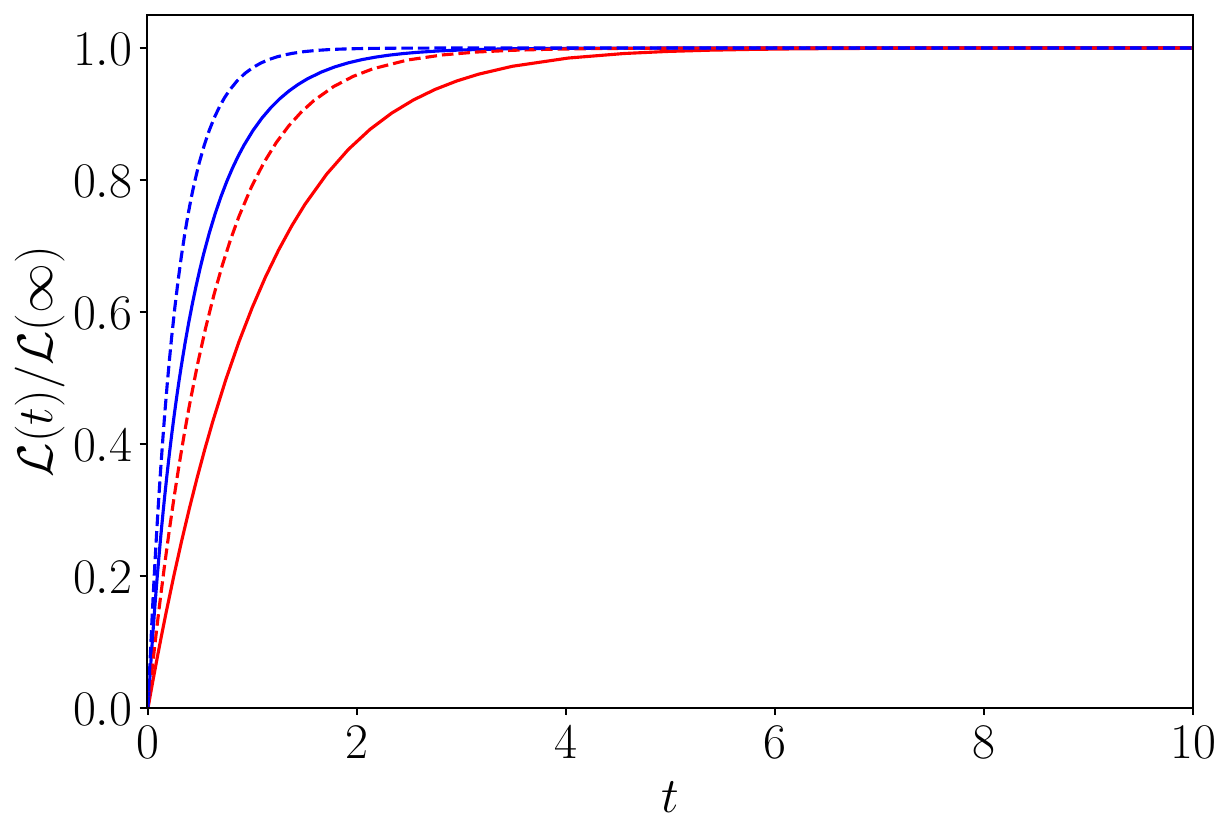}
\caption{KL divergence $D_-^+(t)$ and $D_+^-(t)$ (eq.~\eqref{KL:def1}, top left), statistical velocity $v(t)$  (eq.~\eqref{vT:def}, top right),  statistical velocity ${\mathcal L}(t)$  (eq.~\eqref{LT:def}, bottom left), and degree of completion ${\mathcal L}^f_i(t)/{\mathcal L}^f_i(\infty)$ (bottom right) for a two state system ($N=2$) with  $\beta_-=1.4$, $\beta_+=0.6$, $\epsilon=2$ (full lines) and  $\epsilon=1$ (dashed lines). Notice that  $ D_+^-(t) >D_-^+(t)$ for any $t\ge0$ and $\mathcal L_-^+(\infty)= \mathcal L_+^-(\infty)$. The BB is used to obtain the plots in this figure.}
\label{fig:2states}
\end{figure*}

\subsection{$N=2$ case}
We study first the case of a particle with 2 energy levels (a 1/2 spin particle or a qubit), with stochastic matrix
\begin{displaymath}
\bW=
\begin{pmatrix}
 -w^\up & & w^\down\\
 w^\up & & -w^\down\\
\end{pmatrix}
\end{displaymath}
and with $w^\up=w^\down \exp(-\beta \epsilon)$: it has two eigenvalues $\lambda_{2,0}=0$ and $\lambda_{2,1}=-(w^\up + w_\down)=-w^\up (1+\exp(-\beta \epsilon))$.
Thus we see $|\lambda_{2,1}^+|=|\lambda_{2,1}(\beta_+)|>|\lambda_{2,1}^-|=|\lambda_{2,1}(\beta_-)|$ for any pair of temperatures such that $\beta_+<\beta_-$ ($T_+>T_-$). The decay rate of the heating process is larger than the decay rate of the cooling process: $1/ |\lambda_{2,1}|$ is the only time scale that enters in the convergence towards equilibrium and as such heating is expected to be the faster process.  This is confirmed by the analysis of the quantities introduced above to characterize the convergence towards equilibrium, see fig.~\ref{fig:2states}.
One finds that $D_+^-(t)> D_-^+(t)$ for any $t\ge 0$, indicating that, for a given $t$, the heating process brings the system {\it closer} to its final equilibrium state than the corresponding cooling process.
Furthermore one finds that for the thermodynamic distance, as given by eq.~\eqref{LT:def}, the following inequality holds  $\mathcal L_-^+(t)\ge \mathcal L_+^-(t)$  for any $t\ge 0$, and  is equal for the two processes in the long time regime $\mathcal L_-^+(\infty)= \mathcal L_+^-(\infty)$. 
Therefore such a distance is independent of the thermodynamic direction, as one would expect for a metric measuring distances between equilibrium states. This result for $N=2$  is derived in appendix \ref{symm:N2} and is in accordance with the findings of \cite{Ibanez2024} for the harmonic oscillator. 
Thus, one finds that for the degree of completion, defined as  $\mathcal L_i^f(t)/ \mathcal L_i^f(\infty)$ the following equality holds
\begin{equation}
\mathcal L_-^+(t)/ \mathcal L_-^+(\infty)\ge \mathcal L_+^-(t)/\mathcal L_+^-(\infty)
\end{equation} 
confirming the prediction based on the bare analysis of $\lambda_{2,1}^\pm$ that heating is faster than cooling.

As far as the statistical velocity is concerned, eq.~\eqref{vT:def}, the order relation $v_-^+(t)$ and $v_+^-(t)$  requires a more detailed analysis.

As anticipated above at $t=0$ one finds $v_-^+(0)> v_+^-(0)$, with the same inequality holding at short time, i.e., the heating process is initially the faster one, however at intermediate and large $t$ the inequality is inverted $v_+^-(t)> v_-^+(t)$, see fig.~\ref{fig:2states}.  This can be understood by comparing  the behaviour of $v_i^f(t)$ with $D_i^f(t)$ and $\mathcal L_i^f(t)$ in fig.~\ref{fig:2states} and by  noticing that at intermediate and large time the heating process is {\it closer} to the final equilibrium state. Given that $v(t)$ is related to the rate of change of the instantaneous KL divergence, see eqs.~\eqref{vT:def} and \eqref{DL:I}, we conclude that  statistical velocity decreases upon completion of the equilibration process. 
This is compatible with the inequality $|\lambda_{2,1}^+|>|\lambda_{2,1}^-|$ and with the fact that in the long time regime $v(t)\sim \E^{\lambda_{N,1}t}$.
As discussed above the inequality at initial time $v_-^+(0)> v_+^-(0)$ cannot be proved in general, however a proof for the $N=2$ case is provided in appendix \ref{symm:N2}.
\begin{mycomments}
\alb{ see 2states.nb}
\end{mycomments}

\subsection{$N>2$ case}

\begin{figure}
\center
\includegraphics[width=8cm]{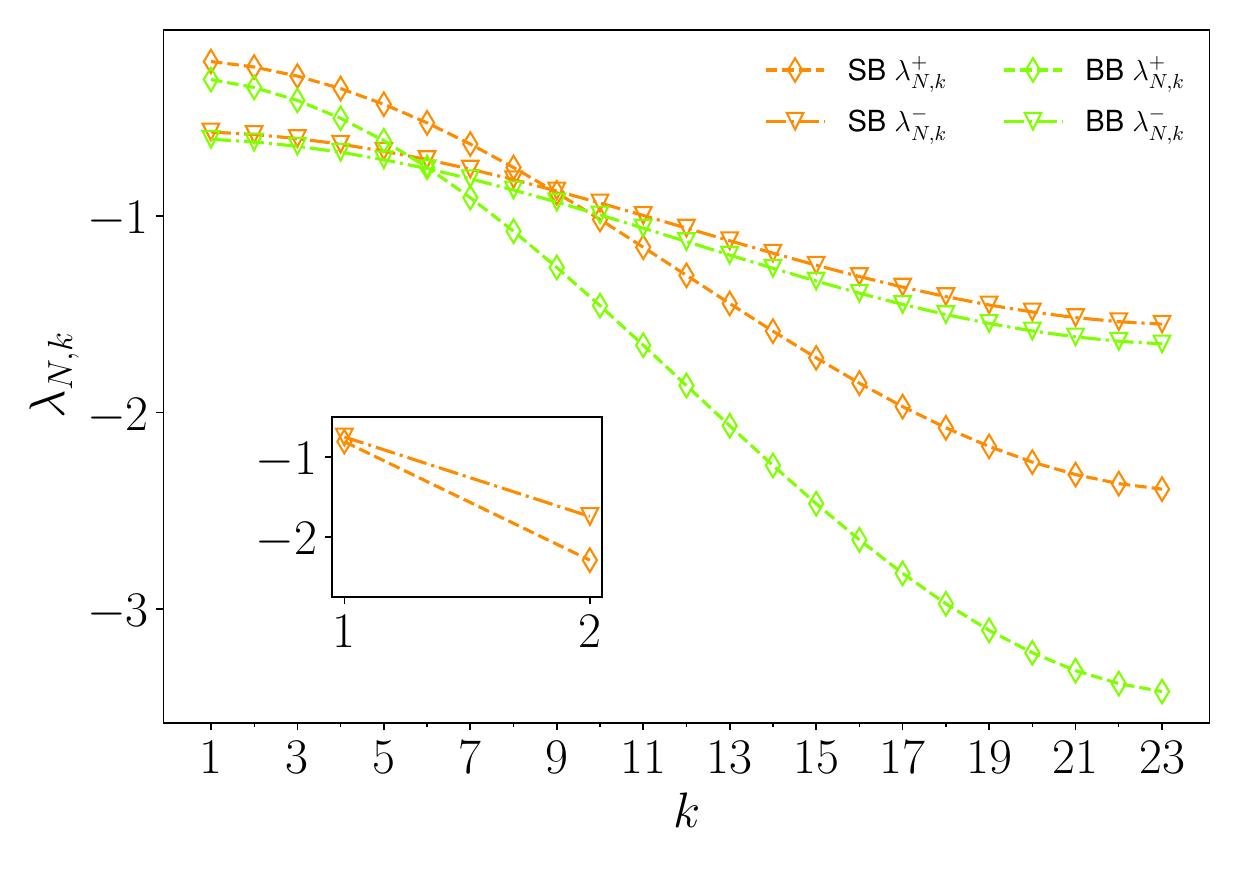}
\includegraphics[width=8cm]{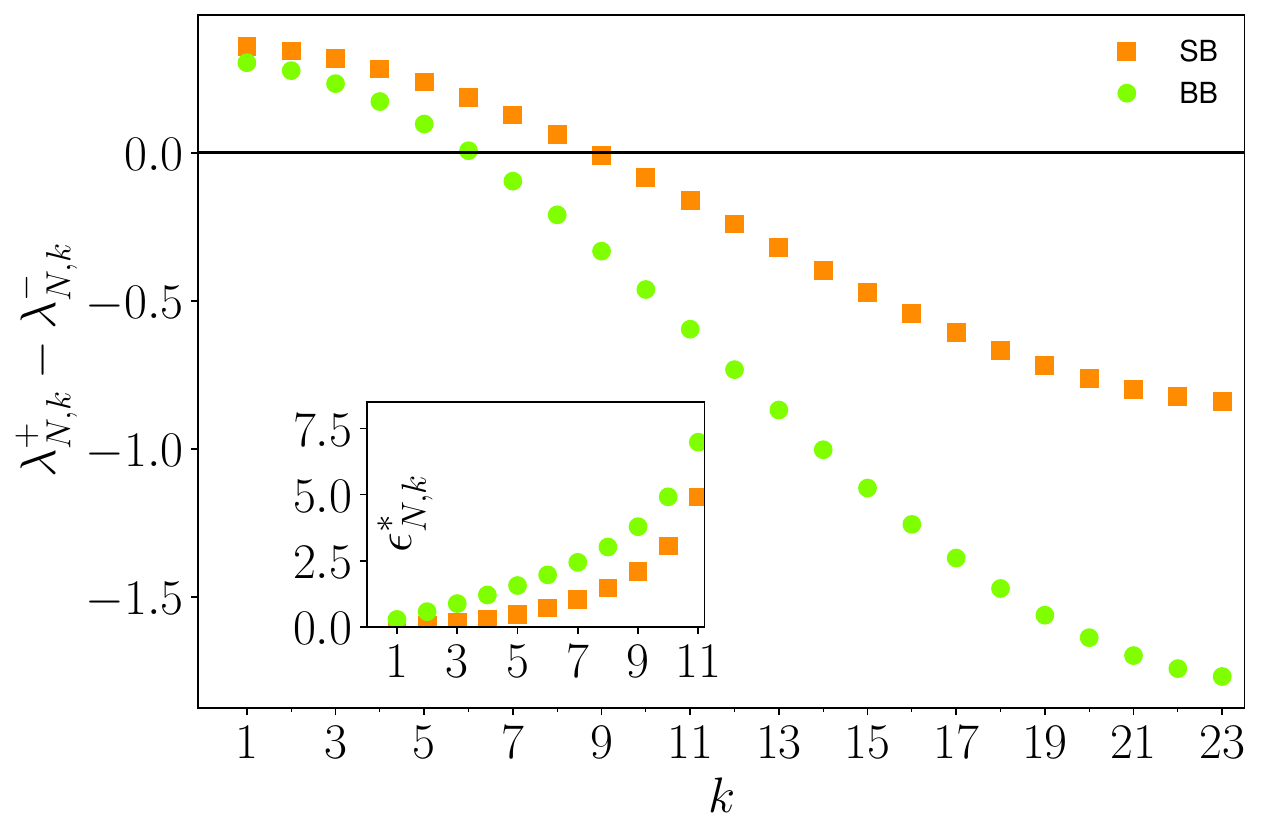}
\includegraphics[width=8cm]{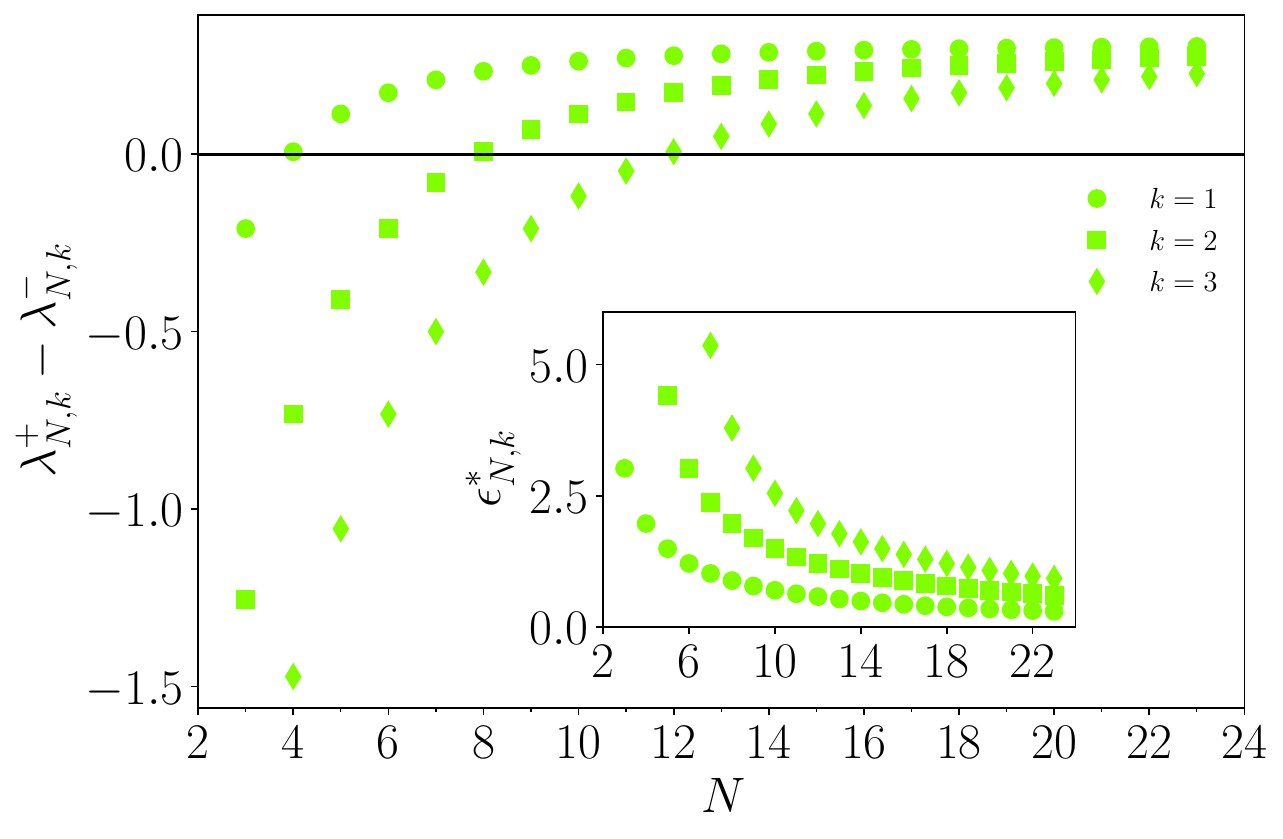}
\caption{First panel: stochastic matrix eigenvalues as a function of the index $k$ for the system with equispaced energy levels, with $\epsilon=2$, $\beta_-=1.4$, $\beta_+=0.6$ and $N=24$ states (main panel), $\epsilon=1$ and $N=3$ (inset), for the two different types of baths SB and BB. Second panel: difference between the eigenvalues of the stochastic matrices $\bW^+$ and $\bW^-$  for the two types of baths and $N=24$. One finds that the inequality $\lambda_{N,k}^+ > \lambda_{N,k}^-$ holds for $1\le k\le 8$ (SB) and for $1\le k\le 6$ (BB). Inset: threshold energy spacing $\epsilon^*_{N,k}$ as a function of $k$ and for fixed $N=24$, see eqs.~\eqref{eq:condition}  and \eqref{epsN:star:Bos}  for the two different types of baths, SB and BB respectively.   Third panel: Difference of the first 3 eigenvalues of the stochastic matrices $\bW^+$ and $\bW^-$ as functions of the system number of states $N$.   Inset: threshold gap as a function of $N$ and for fixed $k=1$, 2, 3 as given by the solution of the equality in eq.~\eqref{epsN:star:Bos} for the  BB. One sees that the values of $\epsilon^*_{N,k}$ as a function of $N$ is a decreasing function of $N$.  }
\label{fig:eigenval}
\end{figure}The situation becomes more interesting if one considers 3 or more states. For the general case with $N$ states the characteristic polynomial of the stochastic matrix \eqref{W:eq} can be obtained exactly, as well as the expression of the single eigenvalues $\lambda_{N,k}$ that reads
\begin{eqnarray}
\lambda_{N,k}&=& -(w^\up + w_\down)+(-1)^{m(k,N)} g_{N,k} \sqrt{w^\up w_\down}
\label{lambda:N}
\end{eqnarray}
with
$$ 
m(k,N)=
\begin{cases}
2, \, \mathrm{if} \, k=1,\dots \left\lfloor \frac{N}{2} \right\rfloor \\
1, \, \mathrm{if} \, k=\left\lfloor \frac{N}{2} \right\rfloor+1 ,\dots N-1\\
\end{cases}
$$
and  where $g_{N,k}$ is a decreasing function of $k$ for  $k\le  \left\lfloor N/2 \right\rfloor$ ,
\begin{equation}
g_{N,k}=\frac{2}{\sqrt{ 1 + \tan^2\p{\frac { k\pi} N}}}
\end{equation} 
 see appendix \ref{eigenv:app} for the derivation.
In particular we want to study the second slowest decay rate in the equilibration processes, namely  
the second largest eigenvalues that  reads $\lambda_{N,1}= -(w^\up + w_\down)+ g_{N,1} \sqrt{w^\up w_\down}$, with 
$1\le g_{N,1}<2$ for $N>2$ and $g_{N,1}\to 2 $ in the limit $N\to \infty$.
We now investigate whether and under which conditions the second slowest decay rate for the cooling process can become larger than the corresponding rate for the heating process. 
In the following we treat the inverse temperatures $\beta_-$ and $\beta_+$ as parameters, and the energy gap $\epsilon$ as a free variable and consider first the SB, the results for the BB turning out to be qualitatively identical.
%One finds that the value $\epsilon^*_N$ for which the second largest eigenvalues satisfy   $|\lambda_{N,1}^+|\le |\lambda_{N,1}^-|$  is solution of  the trascendental equation
%\begin{equation}
%\E^{\epsilon^*_N (\beta_+-\beta_-)/2} =g_{N,1} \E^{ \epsilon^*_N \beta_+/2}-1,
%\label{epsN:star}
%\end{equation} 
One finds that the condition  $|\lambda_{N,1}^+|\le |\lambda_{N,1}^-|$ is equivalent to 
\begin{equation}
g_{N,1} \E^{(\beta_+ + \beta_-) \epsilon/2} \ge   \E^{\beta_+ \epsilon/2} +\E^{\beta_-  \epsilon/2}
\label{eq:condition}
\end{equation} 
For $N=2$ this condition is never satisfied for a non--vanishing $\epsilon$ as $g_{2,1}=0$, while above $N=2$ it is always possible to find a combination of parameters for which it holds.
The functions on the two sides of equation (\ref{eq:condition}) take both positive values, and are both increasing, with the left hand side smaller at small $\epsilon$ but increasing faster than the right hand side, thus for some value of the gap $\epsilon_{N,1}^*$ the two sides of eq.~(\ref{eq:condition}) will become equal, and for $\epsilon> \epsilon_{N,1}^*$ the strict inequality holds, resulting in  $|\lambda_{N,1}^+|< |\lambda_{N,1}^-|$ holds, i.e. the second slowest decay rate for cooling is faster than the corresponding rate for heating. 

We see that for fixed $\beta_+$ and  $\beta_-$, since $g_{N,1}$ increases with $N$, the value of $\epsilon^*_{N,1}$ decreases as $N$ increases. 
In particular as $N\to \infty$, $g_{\infty,1}=2$, and one immediately concludes from eq.~\eqref{eq:condition} that $\epsilon*_\infty=0$, thus $|\lambda_{\infty,1}^+|< |\lambda_{\infty,1}^-|$ for any $\epsilon> 0$. 
Conversely for any arbitrary large but finite $N$, one has  $\epsilon^*_{N,1}>0$, and one can always choose an $\epsilon$ small enough such that  $\epsilon^*_{N,1}>\epsilon>0$. 

The difference between  $\lambda_{N,1}^+$ and $\lambda_{N,1}^-$ vanishes as $\epsilon\to 0$, and one can prove that up to first order in $\epsilon$ the statistical length for $t \to \infty$ is symmetric $\mathcal L_-^+(\infty)\simeq \mathcal L_+^-(\infty)$ with the difference being of order $O(\epsilon^2)$, see appendix \ref{symm:NN} and eq.~\eqref{eq:L:small:eps} in particular.
While the choice $\epsilon=0$ is meaningless, one can regard the classical harmonic oscillator considered in \cite{Ibanez2024}, as the limiting case of the present system with discrete energy levels, in the limit of large $N$ and vanishing $\epsilon$. 
As such, the result of the present paper in this limit is in accordance with that of  \cite{Ibanez2024},  namely that the statistical length is symmetric for a classical harmonic oscillator when $t \to \infty$.

The parallelism finishes here as an harmonic oscillator with mass $m$, proper frequency $\omega_0$ and friction $\gamma$ only has one or two characteristic times describing the decay to equilibrium, which are the inverse of the frequencies $\gamma/m$ or $\gamma/m\pm\sqrt{\gamma^2/m^2 -4\omega_0^2 }$ depending on whether the oscillator is underdamped or not. In any case these characteristic times do not depend on the temperature, and are the same for the heating and cooling process. In other words for a Brownian particle in a harmonic potential the temperature only sets the size of the fluctuations, but not the decay rates towards equilibrium.

On the contrary the discrete model retains $N$ (with $N$ possibly large) time scales, that do depend on the bath temperature.

For higher order eigenvalues
 $\lambda_{N,k}$ (eq.~\eqref{lambda:N}) with $1<k\le\left\lfloor \frac{N}{2} \right\rfloor $ 
an inequality  similar to eq.~\eqref{eq:condition} can be found by replacing $g_{N,1}$ with  $g_{N,k}$, 
 and thus for each set of parameters in the stochastic matrix, one find a threshold value for the gap $\epsilon^*_{N,k}$ for which $|\lambda_{N,k}^+|\le |\lambda_{N,k}^-|$ if $\epsilon> \epsilon^*_k$. 
Given that $g_{N,k}$ is a decreasing function of $k$ (as long as $k\le \left\lfloor \frac{N}{2} \right\rfloor  $), one also has  $ \epsilon^*_{N,k+1}>\epsilon^*_{N,k}$: thus once one fixes all the parameters and the value of $\epsilon$, with  $ \epsilon^*_{N,l+1}> \epsilon>\epsilon^*_{N,l}$, the inequality  $|\lambda_{N,k}^+|\le |\lambda_{N,k}^-|$ is satisfied only for the first $l$ eigenvalues.

The above findings are exemplified by the panels of fig.~\ref{fig:eigenval}, where the eigenvalues $\lambda^{\pm}_{N,k}$ are plotted for some choices of the system parameters.

Inspection of this figure confirms that,  for fixed $\beta_+$, $\beta_-$, $\epsilon$, and $N$,  the difference $\lambda_{N,k}^+-\lambda_{N,k}^-$ as a function of $k$ is first positive for small $k$, and then becomes negative for some value of $k$ that depends on the type of bath.
Conversely, for fixed $k$, the difference $\lambda_{N,k}^+-\lambda_{N,k}^-$ as function of $N$ is first negative for small $N$ and then becomes positive as  $N$ increases. Such a difference becomes positive first for eigenvalues with lower index $k$, which is a consequence of the fact that the threshold value $\epsilon^*_{N,k}$ is an increasing function of $k$, being the solution of the equality in eq.~\eqref{eq:condition} with $g_{N,1}$ replaced by $g_{N,k}$.

In the case of BB, eq.~\eqref{eq:condition} becomes 
\begin{equation}
g_{N,k} \p{\E^{(\beta_+ + \beta_-) \epsilon/2} +1} \ge 2 \p{ \E^{\beta_+ \epsilon/2} +\E^{\beta_-  \epsilon/2}}
\label{epsN:star:Bos}
\end{equation} 
This gives a different value of the threshold gap $\epsilon^*_{N,k}$, but the qualitative behaviour of the eigenvalues being identical to that obtained for the SB.
\begin{mycomments}
\alb{ see study\_lambda\_12\_05\_2024.nb} 
\end{mycomments}

\begin{figure*}[h]
\center
\psfrag{ }[ct][ct][1.]{ }
\includegraphics[width=0.49\textwidth]{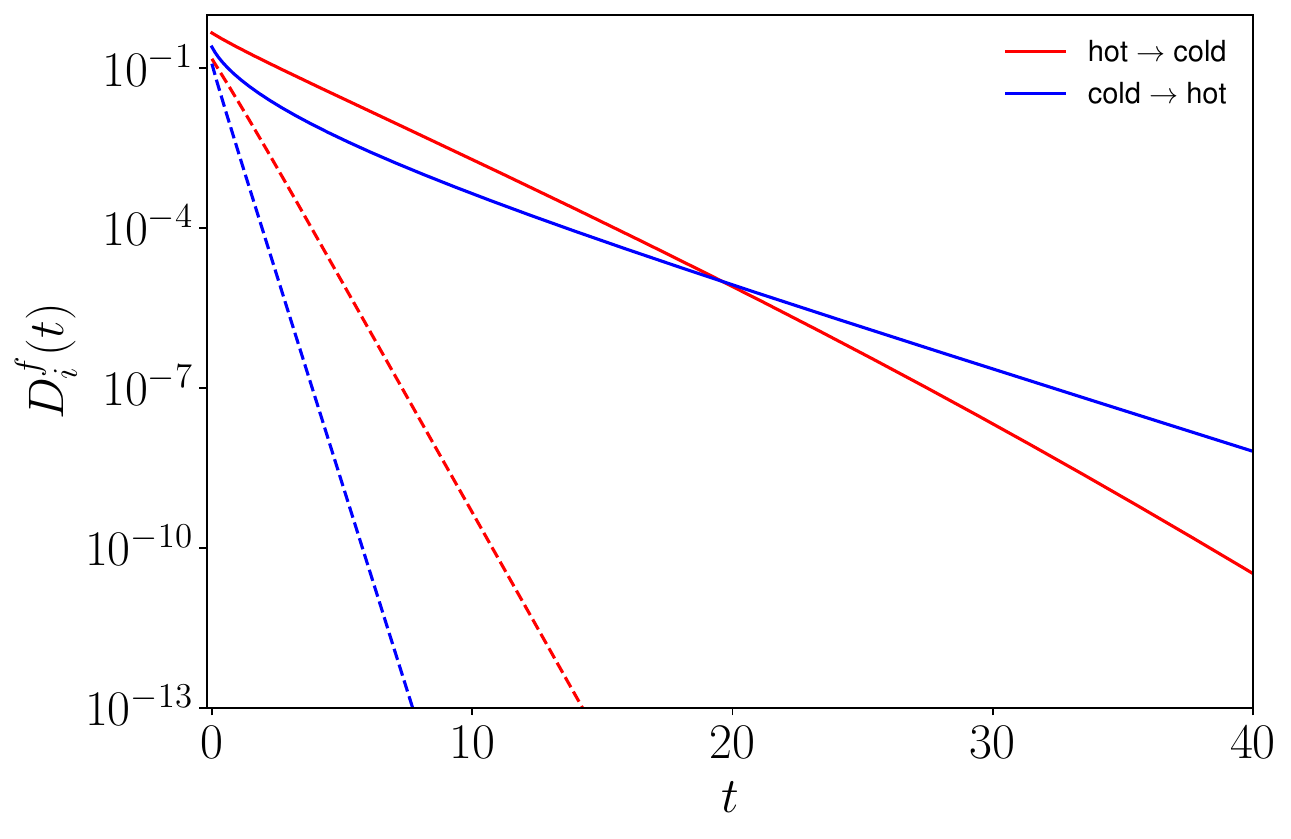}
\includegraphics[width=0.49\textwidth]{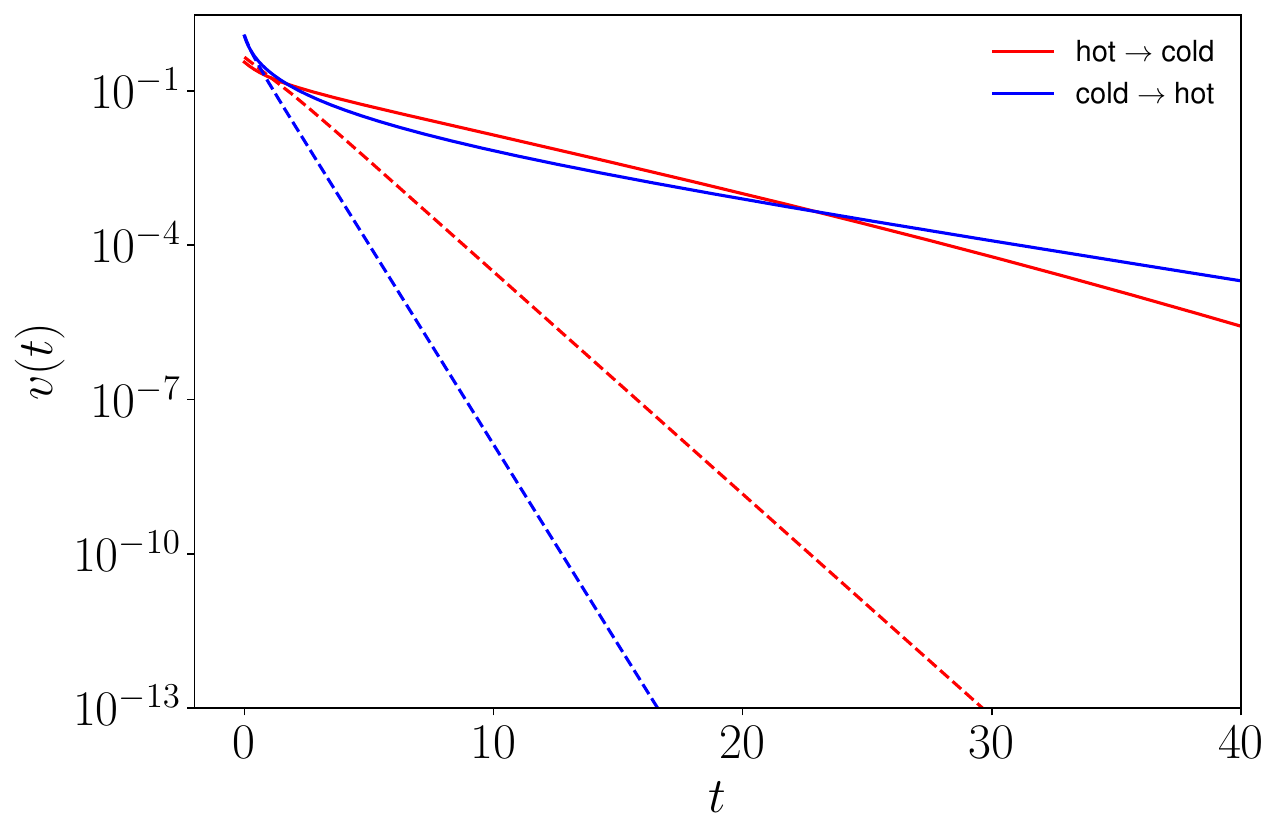}
\includegraphics[width=0.49\textwidth]{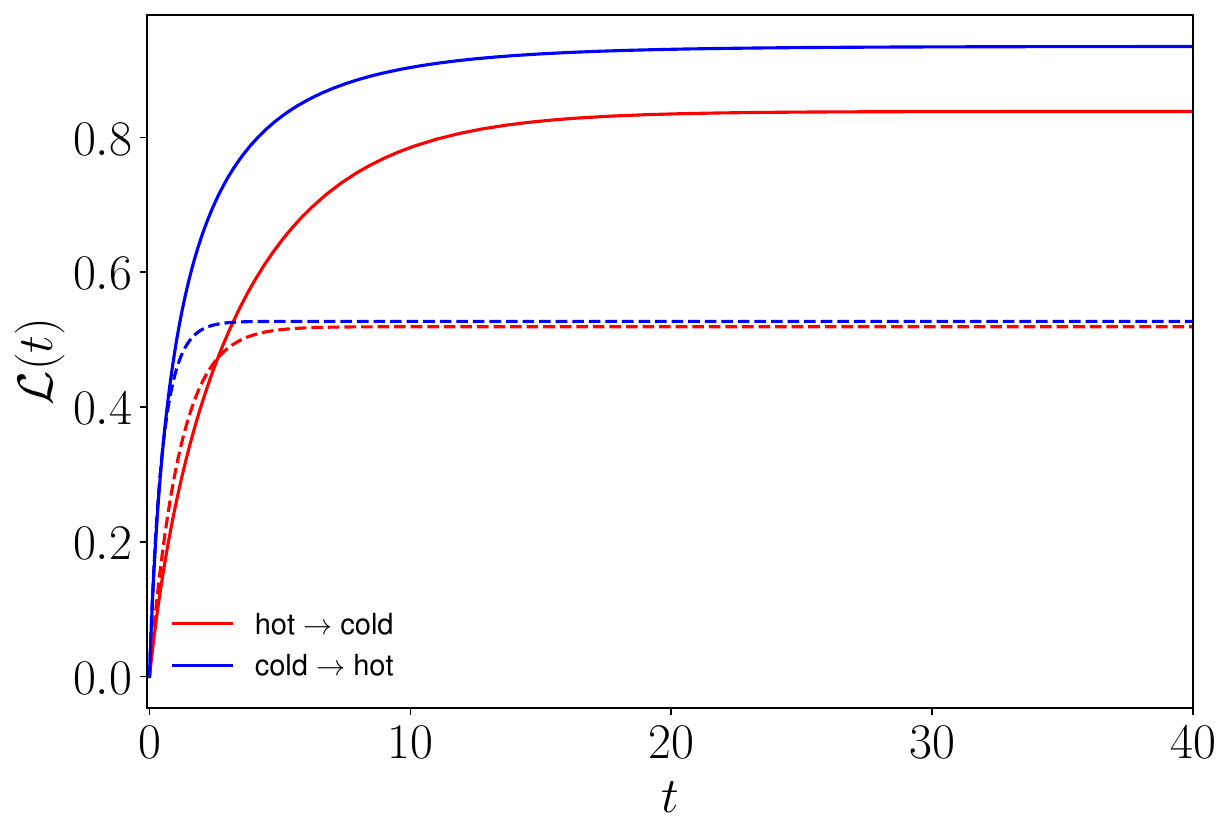}
\includegraphics[width=0.49\textwidth]{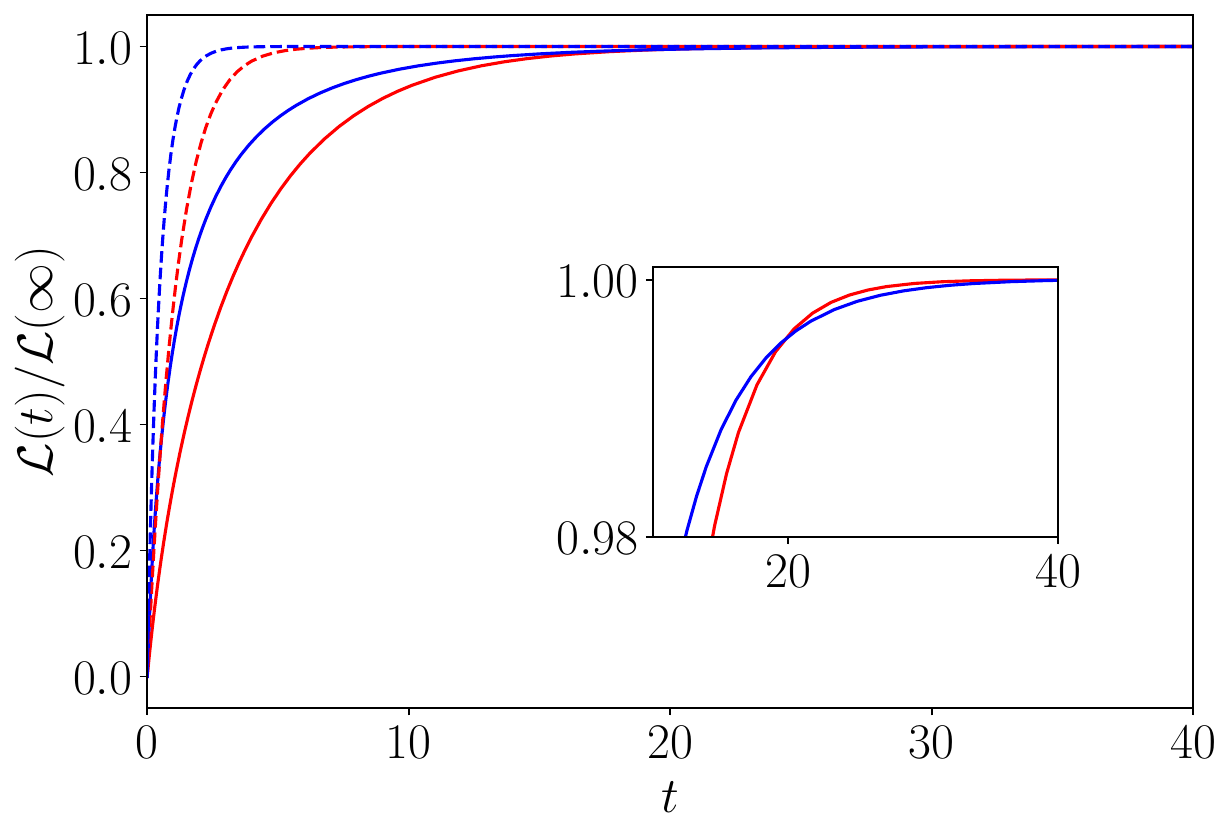}
\caption{KL divergence, statistical velocity, statistical length and completion rate as functions of $t$, with $\epsilon=1$, $\beta_-=1.4$, $\beta_+=0.6$ and $N=24$ states (full line), $N=3$ dashed line (same parameter as fig.~\eqref{fig:eigenval}) and BB. For $N=24$ and in the intermediate and long time regime one clearly sees the inversion of the convergence rate, with the cooling process becoming the one converging faster to equilibrium. In the log-linear plots of $D(t)$ and $v(t)$ the slopes in the linear regions are $2 \lambda_1$ and  $\lambda_1$, respectively (lines not shown), as predicted from long--time analysis discussed in appendix \ref{appendix:two}.}
\label{fig:L:t}
\end{figure*}

Having analyzed the behaviour of the eigenvalues $\lambda_{N,k}^\pm$ we can now study the convergence to equilibrium of the heating and the cooling processes, and in particular the KL divergence, the statistical velocity and the statistical length.
A plot of such quantities is shown in fig.~\eqref{fig:L:t}.
In particular, for $N=3$ the chosen value of $\epsilon$ is too small to obtain the inversion of the order of the eigenvalues, thus one has 
 $|\lambda_{3,k}^+|> |\lambda_{3,k}^-|$,  $k=1,\, 2$, the KL relative entropy converges faster to its equilibrium value for the heating process, the same is found for statistical velocity, with  $v(t)$ going to zero faster for the heating process, and the statistical length satisfying at all times  $\mathcal L_-^+(t)\ge \mathcal L_+^-(t)$. When $N$ is large ($N=24$ in fig.~\ref{fig:L:t}) a completely different behaviour appears in the intermediate and long time regime: the KL relative entropy converges faster to its equilibrium value for the {\it cooling} process, thus the statistical velocity $v(t)$ goes to zero faster than the one for the {\it heating} process.
We remind the reader that the statistical velocity is indeed related to the rate of change of the instantaneous DL divergence, eqs.~\eqref{vT:def} and \eqref{DL:I}.
In other words, since the {\it cooling} process is completed earlier, its pace, as measured by $v(t)$ slows down. Conversely in the long time limit the {\it heating} process still proceeds at higher pace, as the KL divergence is further away from the equilibrium value $D_-^+(t) > D_+^-(t)$ and thus the system retains a higher velocity resulting in $v_-^+(t) > v_+^-(t)$.  Mathematically this is a consequence of the inversion of the order of the eigenvalues, i.e.  $|\lambda_{N,k}^+|< |\lambda_{N,k}^-|$, with the cooling process becoming the one with the shortest relaxation time and with the heating process taking more time to fill up the high energy levels.
We remind the reader that in the long time limit we have proved $D(\bp_i^f(t)|| \bpeq_f) \sim\exp(2 \lambda^f_1 t)$ (eq.~\eqref{D:longt}) and a similar result has been obtained for $I(t)=v^2(t)$. The physical intuition behind these results is that, given the initial states as depicted in fig.~\ref{sistema:fig}, for $\epsilon$ large enough the {\it cooling} process becomes the faster one with the higher energy levels being emptied at a greater speed, while in the {\it heating} process it takes  longer time for the dynamic to explore the phase space and in particular the states with higher energy.  

While for the statistical length the equality   $\mathcal L_-^+(t)\ge \mathcal L_+^-(t)$ still holds at any time, the change at intermediate time scale is reflected in a different behaviour of the completion rate, where the inequality now holding in the opposite direction 
  $\mathcal L_-^+(t)/\mathcal L_-^+(\infty) \le \mathcal L_+^-(t)/\mathcal L_+^-(\infty)$, signaling that the cooling process is completed first, see inset in the fourth panel of fig.~\ref{fig:L:t}.

\begin{figure}[h]
\center
\psfrag{ }[ct][ct][1.]{ }
\includegraphics[width=0.49\textwidth]{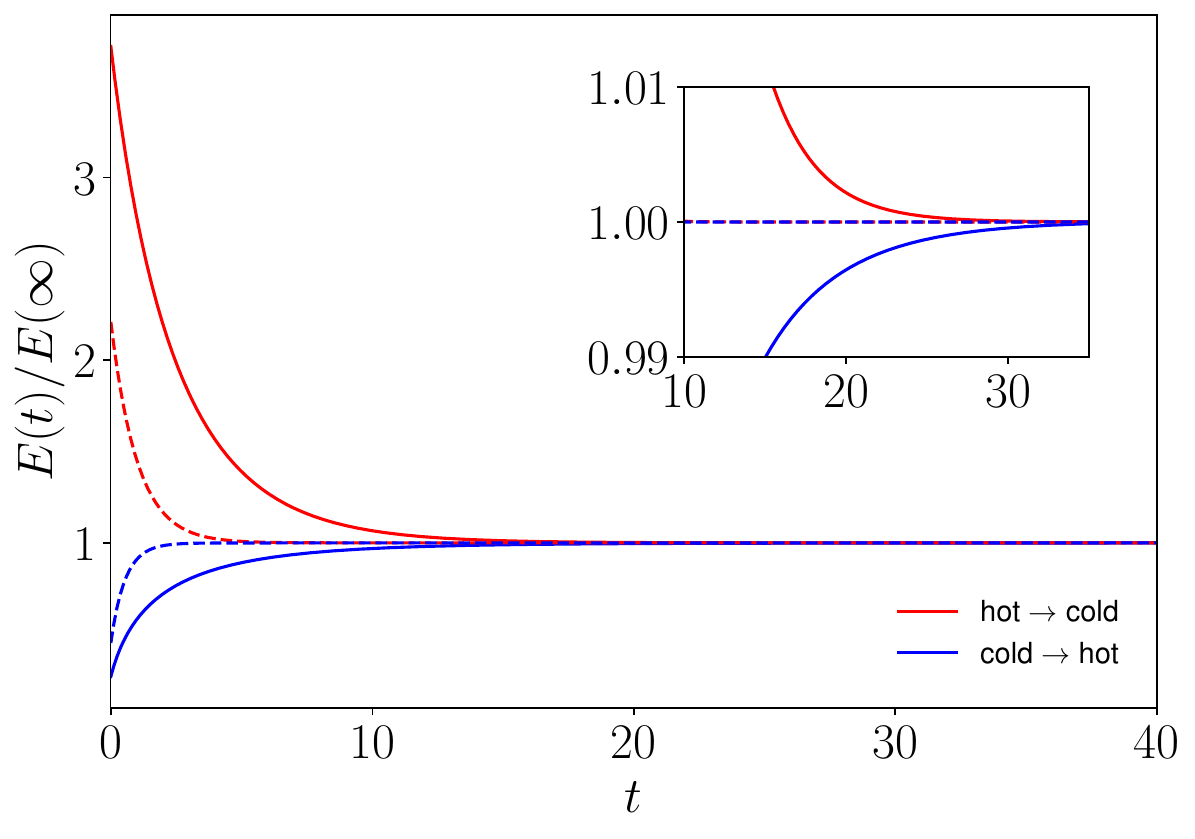}
\includegraphics[width=0.49\textwidth]{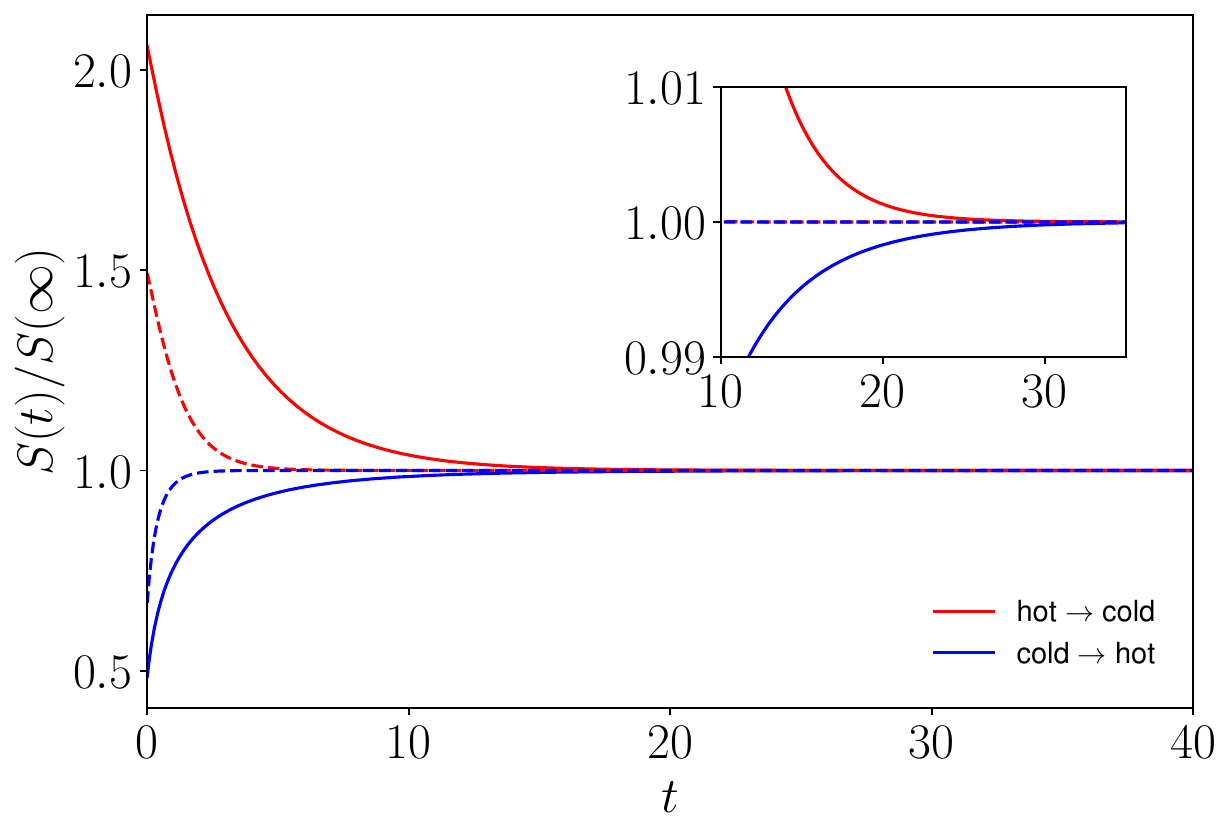}
\caption{Rescaled average energy and stochastic entropy as functions of the time. The values of the parameters are the same as in fig.~\ref{fig:L:t}.  One notices that in the intermediate and long time regime the cooling process converges faster.}
\label{E:S:t}
\end{figure}

The above results are confirmed by the analysys of the average energy $E(t)=\sum_k p_k(t) \epsilon_k$ and of the  average stochastic entropy $S(t)=-\sum_k p_k(t)\log p_k(t)$: provided the energy gaps are large enough, in the intermediate and long time regime  both $E(t)$ and $S(t)$ converge to their equilibrium values faster for the {\it cooling} process than for the heating one, see fig.~\ref{E:S:t}.

We conclude this section by noticing that in the long time limit we always find the strict inequality $\mathcal L_-^+(\infty) >\mathcal  L_+^-(\infty)$ for finite $\epsilon$ and $N>2$ at variance with the findings of \cite{Ibanez2024} for a classical harmonic oscillator, where the two lengths were found to be equal.

\begin{mycomments}
\alb{in eigenvalues.py and eigenvalues\_2states.py I have added the calculation and the plots of the energy and stochastic entropy as a function  of $t$. Indeed the KL div. is the difference between the out of eq. free energy and the equilibrium final one, see notes\_spring\_2024.pdf}
\end{mycomments}

\section{Thermalization in an Ising model}
\label{deg:sec}

\begin{figure}[h]
\center
\includegraphics[width=8cm]{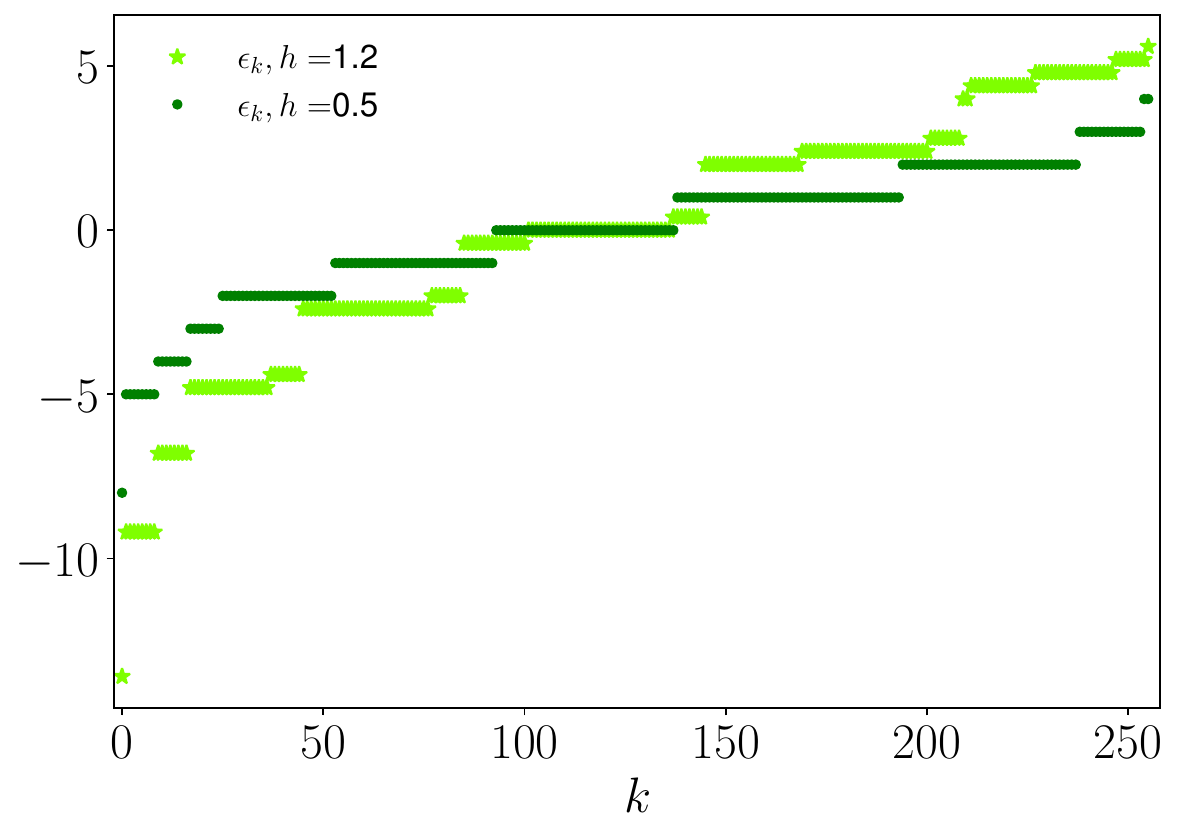}
\caption{Energy eigenvalues of the model \eqref{H:spin} with $M=8$ spins,  $J=0.5$ and for two different values of $h$.}
\label{energy:fig}
\end{figure}

The stochastic matrix in eq.~\eqref{W:eq} describes a simple topology of a linear network of states, where each state is only connected to two nearest neighbour states.
In order to consider a more complex topology, where each state is connected to several states, we study the stochastic dynamics of a $M$-spin system, with Hamiltonian
\begin{equation}
H=-\sum_{j=0}^{M-1} (J\sigma_j \sigma_{j+1} + h \sigma_j ).
\label{H:spin}
\end{equation} 
In order to investigate the thermalization of such a model, one can build on the findings of the previous section: upon increasing of the energy gap $\epsilon$ in a model with equispaced energy levels  there is a crossover from a regime where where the heating process is faster than the cooling one on all the time scales to a regime where the cooling process becomes faster,  at least for the first $l$ slowest rates.
In the model \eqref{H:spin}, increasing $h$ would increase the gaps between the different energy shells of the model in eq.~\eqref{H:spin} and reduce the degeneracy of the energy eigenvalues.
This is exemplified in fig.~\ref{energy:fig}, where the eigenenergies $\epsilon_k$ of \eqref{H:spin} are plotted as a function of $k$ for two different values of $h$.

We thus consider the dynamics generated by the stochastic matrix with elements \eqref{omega:Bos}, connecting two states differing by only one spin orientation, i.e., given a microscopic state $\pg{\sigma_j}$, only one spin is allowed to flip in a single stochastic jump. With this choice, each microscopic state is connected to $M$ other states.

\begin{figure}[h]
\center
\psfrag{ }[ct][ct][1.]{ }
\includegraphics[width=8cm]{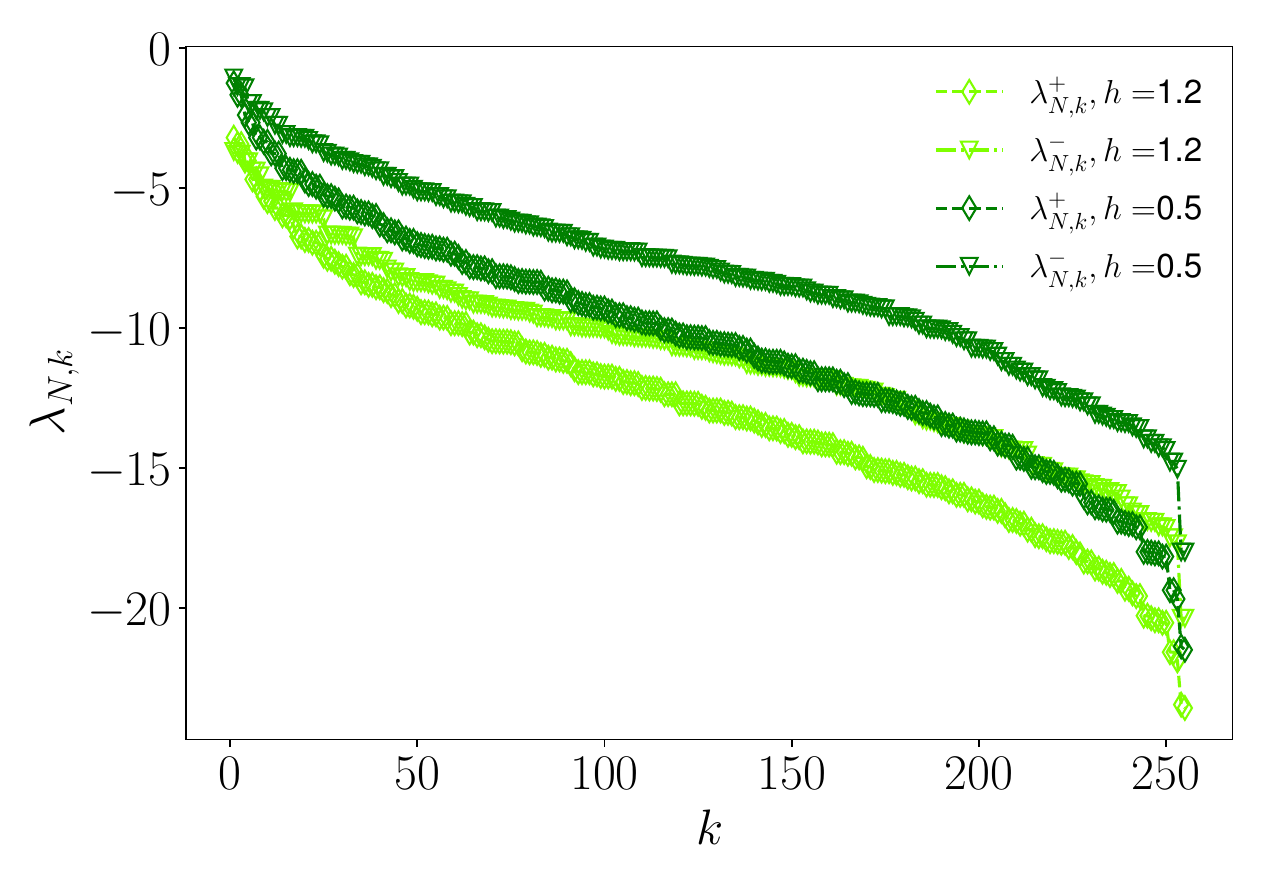}
\includegraphics[width=8cm]{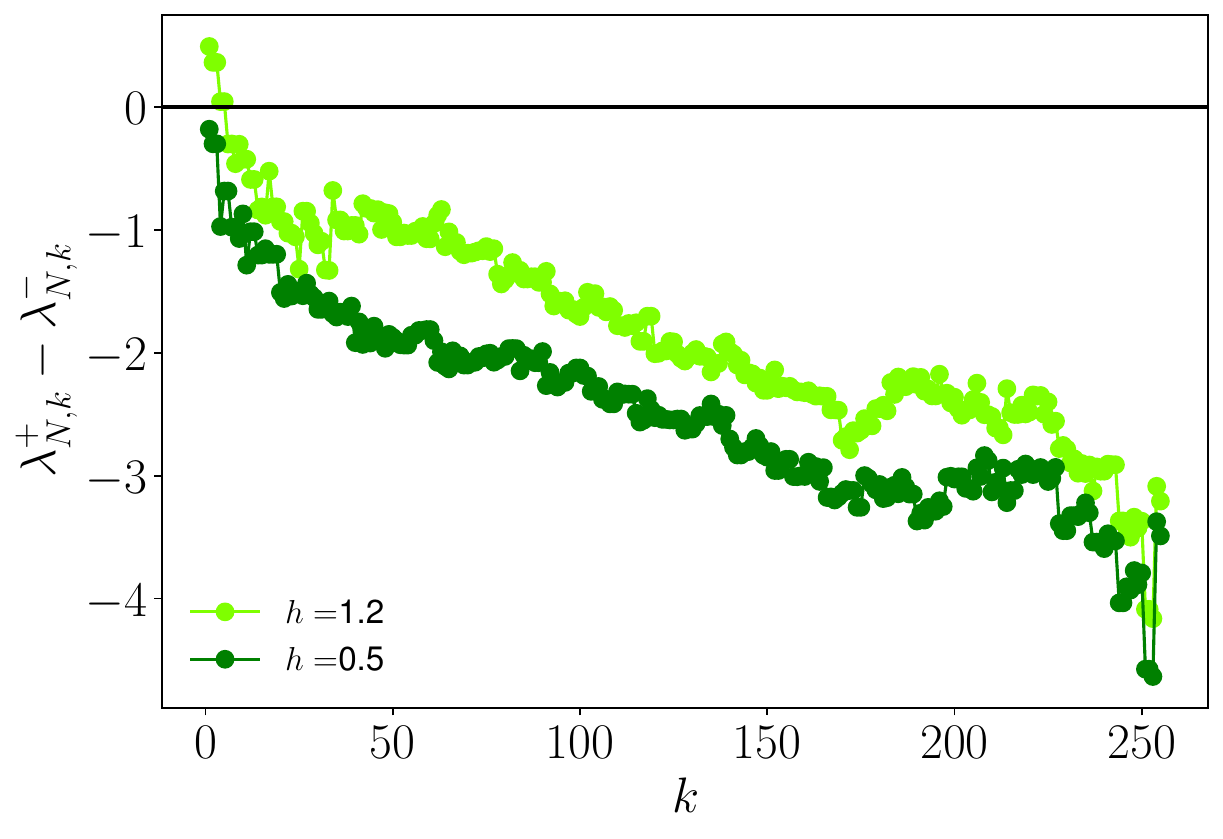}
\caption{Top panel: stochastic matrix eigenvalues as a function of the index $k$ for the spin system with Hamiltonian as given in eq.~\eqref{H:spin}, with $M=8$ spins, $J=0.5$, $\beta_-=1.4$, $\beta_+=0.9$, for two values of the external field $h$ and BB rates, eq.~\eqref{omega:Bos}. Bottom panel: difference between the eigenvalues of the stochastic matrix with inverse temperature $\beta_+$ and $\beta_-$.  }
\label{fig:eigenval_Nspins}
\end{figure}

The eigenvalues $\lambda_{N,k}^\pm$ and their difference are plotted in fig.~\ref{fig:eigenval_Nspins} as a function of $k$, with $N=2^M$. We see that by increasing $h$ the eigenvalues show inversion of the order, i.e.  $|\lambda_{N,k}^+|< |\lambda_{N,k}^-|$, at least for a few initial values of $k$, corresponding to the longest time scales.

\begin{figure*}[h]
\center
\psfrag{ }[ct][ct][1.]{ }
\includegraphics[width=8cm]{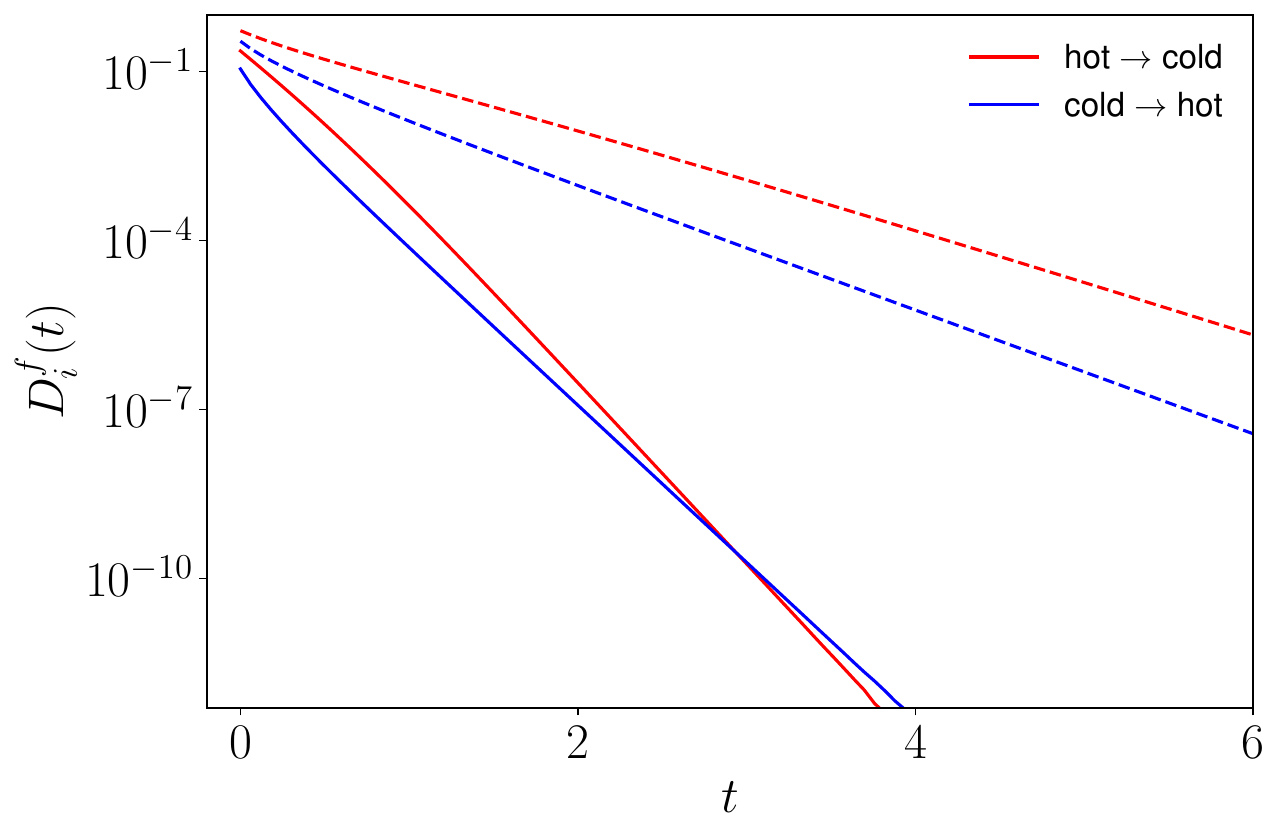}
\includegraphics[width=8cm]{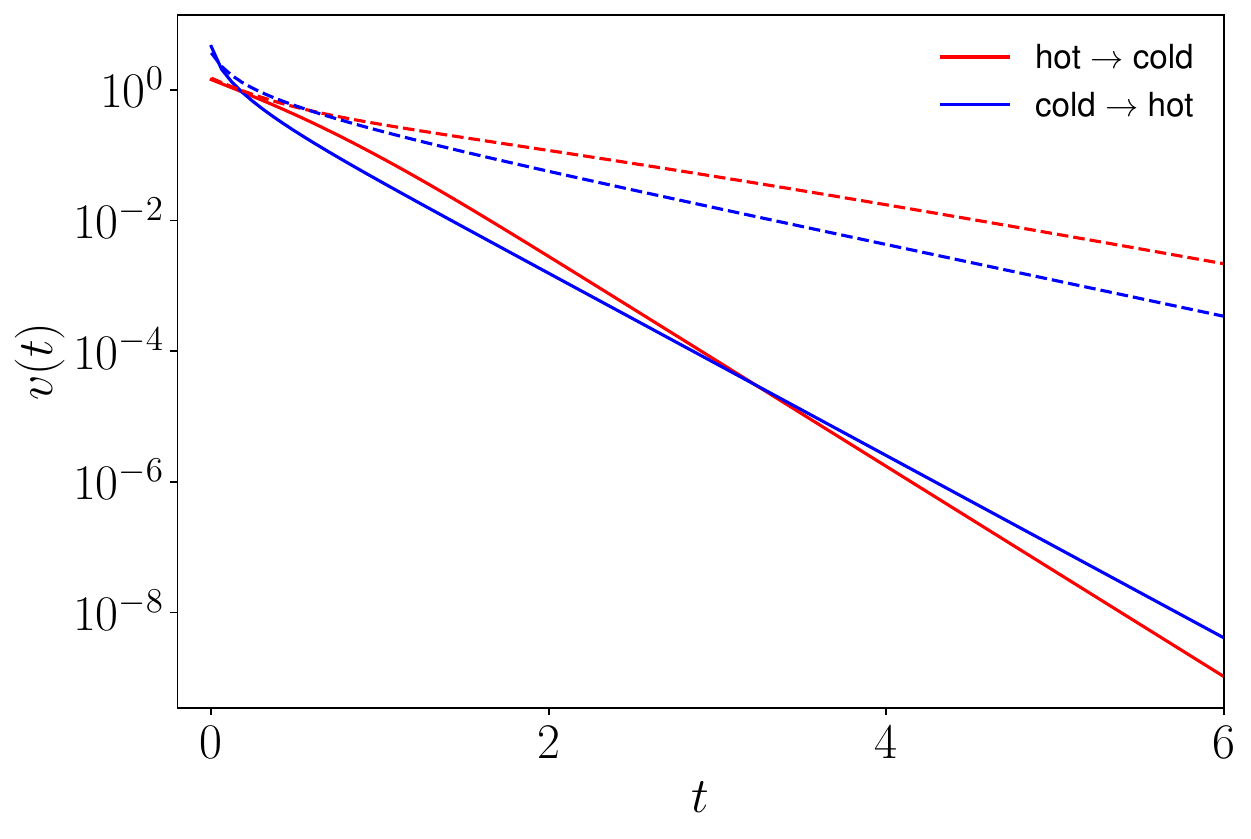}
\includegraphics[width=0.49\textwidth]{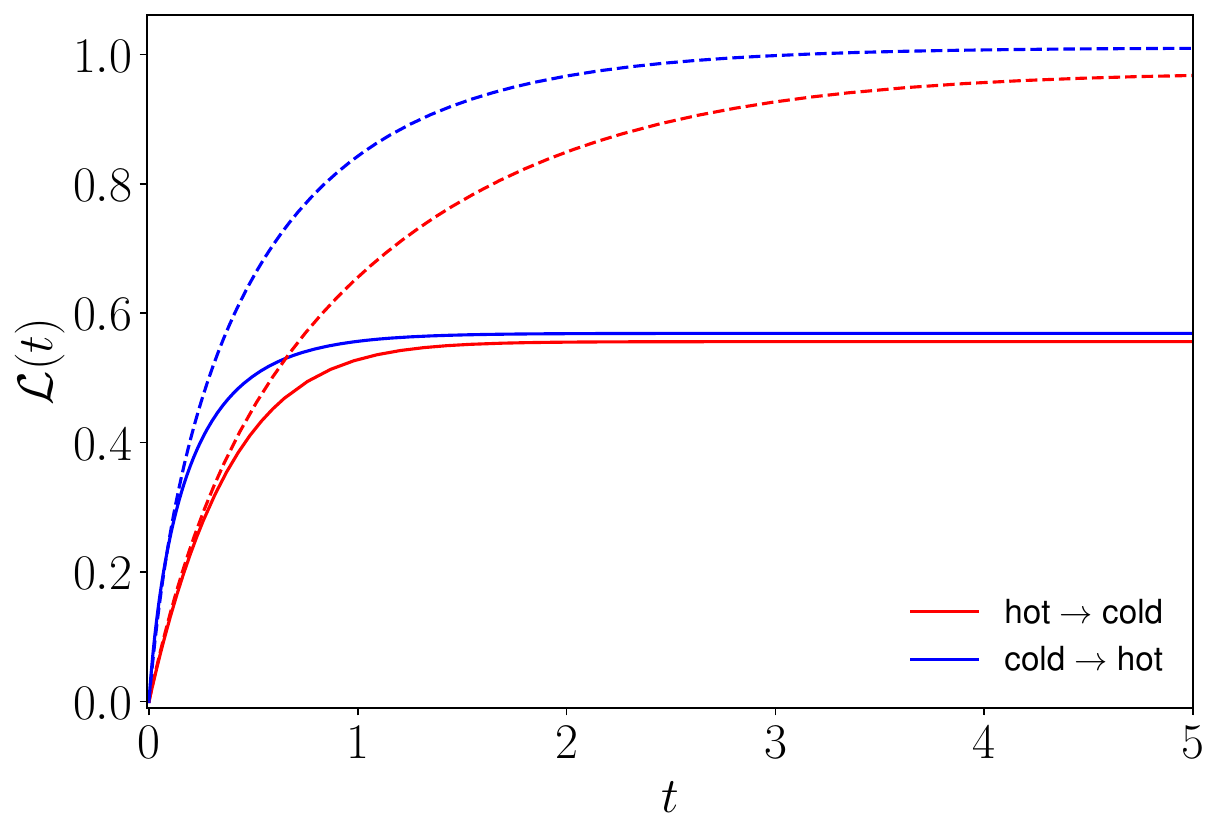}
\includegraphics[width=0.49\textwidth]{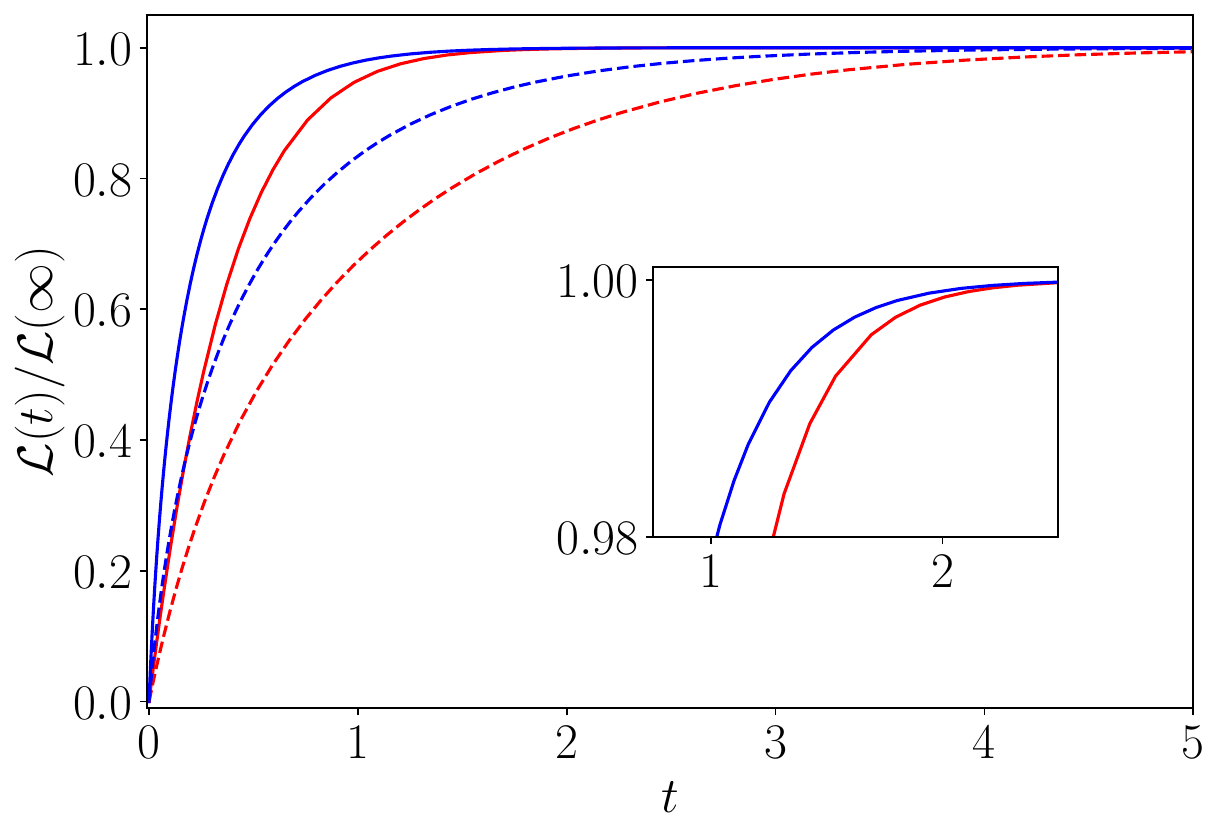}
\caption{KL divergence $D_-^+(t)$ and $D_+^-(t)$ (eq.~\eqref{KL:def1}, top left), statistical velocity $v(t)$  (eq.~\eqref{vT:def}, top right),  statistical velocity ${\mathcal L}(t)$  (eq.~\eqref{LT:def}, bottom left), and degree of completion ${\mathcal L}^f_i(t)/{\mathcal L}^f_i(\infty)$ (bottom right)
 for the spin system in eq.~\eqref{H:spin} with   $M=8$ spins,  $J=0.5$, $\beta_-=1.4$, $\beta_+=0.9$,  $h=1.2$ (full lines),   and $h=0.5$ (dashed lines), BB.  In the log-linear plots of $D(t)$ and $v(t)$ the slopes in the linear regions are $2 \lambda_1$ and  $\lambda_1$, respectively (lines not shown), as predicted from long--time analysis discussed in appendix \ref{appendix:two}.}
\label{fig:D:v:t_Nspins}
\end{figure*}

This is then reflected in the KL divergence and in the statistical velocity, where for relatively small $h$ one finds again that the heating process is the faster to converge to equilibrium, while for larger $h$ the cooling process becomes the faster to converge to equilibrium in the intermediate and long time scale, see fig.~\ref{fig:D:v:t_Nspins}.
The statistical length is found again to be non--symmetric between the two processes, signaling different pathways to equilibrium for the two opposite processes.

\section{Conclusions}
\begin{mycomments}
\alb{Oscillatore armonico un pathway, come lo spin singolo}

\alb{Do \cite{Ibanez2024} and \cite{Ito2020} state explicitly that $\mathcal{L}$ is a metric?
Because our result on $\mathcal{L}^+_-(\infty)\neq \mathcal{L}_+^-(\infty)$ contradict that.
Yes, \cite{Ibanez2024} says that before eq. 3}
\end{mycomments}
Our findings suggest that for physical systems with a detailed model of system-bath interaction, the heating or the cooling process can become faster, depending on the choice of system parameters. In particular choosing large energy gaps leads to slower relaxation when in contact with the hot bath, compared to the case when the equilibration occurs in the cold bath.

In this sense the single spin and the harmonic oscillator seem to represent exceptions and do not exhibit a crossover between these different behaviours. For these two systems the heating is always the faster process, as they are characterized by one or two characteristic times, which do not exhibit inversion of the order of eigenvalues describing convergence to equilibrium, and in the case of the harmonic oscillator are independent of the temperature. As such, the heating and the cooling process for these systems turn out  to be intrinsically symmetric.

The use of the statistical length appears problematic in this context: while it is tempting to base a general formulation for a thermal kinematics theory on such a quantity, it turns out to be intrinsically asymmetric for two opposite simple, yet general, processes. This, in turn, signals that two  opposite stochastic processes can be dominated by inherently different pathways.
Furthermore, if one is interested in a measure of the completion of equilibration of a stochastic process, the KL divergence and the statistical velocity (or the Fisher information) seem to provide the same information starting from the intermediate time range, as they are both dominated by the second largest eigenvalue of the stochastic matrix. Thus in this range, where few slow processes dominates the convergence to equilibrium, there is no specific reason to prefer one indicator over another.

We emphasize that, while throughout this paper we have used the energy gaps as adjustable parameters to tune the inversion of the time scales, decreasing the temperatures $T_+$ and $T_-$ by the same factor is equivalent to increase the size of the gaps appearing in the stochastic matrices.

In the opinion of the author of the present paper the analysis of the entire spectrum of the stochastic matrix (or of the Fokker--Planck operator when relevant) represents the safest and more detailed method to asses whether one process is {\it faster} than the other, as it gives a way of comparing all the time scales in a transparent manner. This approach iii) can then be added up to the two introduced in section \ref{sec:I}. 
Both the KL divergence and the Fisher information (and its related quantities) while useful and widely employed quantities in stochastic thermodynamics,  represent global rather than detailed evaluators of the convergence to equilibrium, depending nontrivially  on both the initial conditions and the stochastic matrix eigenvalues.

While the formalism of the present paper is the one of classical stochastic systems, we believe that the results presented are relevant also in quantum regime,  where the dynamics is described by a quantum master equation and and the energy levels are quantized.
Indeed for a {\it global} quantum master equation with  jump operators connecting  eigenstates of the system Hamiltonian $H$, if the initial state is one of such eigenstates, the dynamics reduces to the one described by the classical master equation considered here, see, e.g, \cite{Adam2021,DeChiara2022}.

Finally, the present theoretical study can be subject to experimental verification by using, e.g., a colloidal particle in a ladder-like potential, as realized with the setup discussed in \cite{Toyabe2010} or \cite{Ciliberto2015}.

\begin{acknowledgments} 
I am grateful to Sascha Wald for interesting discussions at the beginning of this project. I am also grateful to Aljaz Godec for insightful comments, in particular on the convergence of the average energy and stochastic entropy.
\end{acknowledgments}

\appendix
\section{Additional results on KL divergence and Fisher Information}
\label{appendix:one}

We want to prove the inequality 
\begin{eqnarray}
&& D_+^-(0)- D_-^+(0)=\nonumber\\
&&=\sum_k \peq_{+,k}\log \p{\frac  {\peq_{+,k}}{ \peq_{-,k}} }- \peq_{-,k}\log \p{\frac  {\peq_{-,k}}{ \peq_{+,k}} }\ge0 \qquad\nonumber \\
&& {\it iff}\, T_+\ge T_-.
\label{D:inequality}
\end{eqnarray} 
We treat the above difference as a function of $\beta_+=1/T_+$ with $\beta_-=1/T_-$ a constant parameter, and introduce 
\begin{eqnarray}
&&f(\beta_+)= D_+^-(0)- D_-^+(0)=\nonumber \\
&&=\sum_k \peq_{+,i}\log \p{\frac  {\peq_{+,k}}{ \peq_{-,k}} }- \peq_{-,k}\log \p{\frac  {\peq_{-,k}}{ \peq_{+,k}} };
\label{fbeta}
\end{eqnarray} 
and notice that
\begin{equation}
f(\beta_+=\beta_-)=0.
\end{equation} 
We also notice that
\begin{eqnarray}
&&\partial_{\beta_+} f(\beta_+)=\sum_k\p{\frac{\peq_{-,k}}{\peq_{+,k}}- \log \frac{\peq_{-,k}}{\peq_{+,k}}}\partial_{\beta_+} \peq_{+,k}\nonumber\\
&&=\sum_k\p{\frac{\peq_{-,k}}{\peq_{+,k}}- \log \frac{\peq_{-,k}}{\peq_{+,k}}}\p{-\epsilon_k- \partial_{\beta_+} \ln Z_+}\peq_{+,k}=\nonumber\\
&&=\sum_k\p{\frac{\peq_{-,k}}{\peq_{+,k}}+ (\beta_- - \beta_+) \epsilon_k -\log \frac{Z_+}{Z_-}} \p{-\epsilon_k + \Epm}\peq_{+,k}\nonumber\\
&&=\Epm-\Emm-(\beta_- - \beta_+) \p{\Esqpm-(\Epm)^2},
\label{der:fbeta}
\end{eqnarray} 
and finally
\begin{eqnarray}
&&\frac{\partial^2}{\partial \beta_+^2} f(\beta_+)=\nonumber\\
&&=\partial_{\beta_+} \Epm+ \Esqpm-(\Epm)^2\nonumber\\
&& \quad+ (\beta_+ -\beta_-) \partial_{\beta_+}\p{\Esqpm-(\Epm)^2}\nonumber\\
&&=(\beta_+ -\beta_-) \partial_{\beta_+}\p{\Esqpm-(\Epm)^2}.
\label{sec:der:fbeta}
\end{eqnarray} 
We now show that the second derivative of $f(\beta_+)$ is always non-negative. By noticing that $\p{\Esqpm-(\Epm)^2}=T_+^2 C_+$, where $C_+$ is the specific heath, we obtain
\begin{equation}
\partial_{\beta_+}\p{\Esqpm-(\Epm)^2}=-\frac{2}{\beta_+^3} C_+ + \frac{1}{\beta_+^2} \partial_{\beta_+}C_+, 
\label{sec:der:fbeta1}
\end{equation} 
which is negative given that $\partial_{\beta_+}C_+<0$ for thermodynamic stability.
Thus from eq.~\eqref{sec:der:fbeta}--\eqref{sec:der:fbeta1} we conclude that
\begin{equation}
\frac{\partial^2}{\partial \beta_+^2} f(\beta_+)\ge0 \qquad {\it iff}\, \beta_+\le \beta_- ,
\end{equation}
the equality holding for equal temperatures.
The last result tells us that $f'(\beta_+)$, as given by eq.~\eqref{der:fbeta}, is an increasing function of $\beta_+$ for  $\beta_+< \beta_-$, and by noticing that  $f'(\beta_+=\beta_-)=0$, we conclude that $f'(\beta_+)\le 0 $ for $\beta_+ \le \beta_-$.

We now consider the function $f(\beta_+)$ as given by eq.~\eqref{fbeta}, and noticing  that $f(\beta_+=\beta_-)=0$ and that   $f(\beta_+)$ is decreasing  for $\beta_+ < \beta_-$, we conclude that $f(\beta_+)\ge 0$  for $\beta_+ \le \beta_-$, which proves eq.~\eqref{D:inequality}. 

It is worth noticing that throughout this paper we are considering finite temperatures, and non-critical systems.

\section{Expansion in eigenfunctions}
\label{appendix:two}
Here we review the expansion of the solution of a master equation in the eigenvectors of the corresponding stochastic matrix, as discussed in, e.g.,  \cite{vanKampen1981}.

Given a stochastic matrix $\bW$ with eigenvalues $\lambda_n$ and eigenvectors $\mathbf{\Phi}_n$, the components of the vector $\bp$ solution to the master equation \eqref{eq:mast} read
\begin{equation}
p_k(t)= \Phi_{0,k}+ \sum_{n=1}^{N-1}  c_n \E^{\lambda_n t } \Phi_{n,k},
\label{pi:eigenve}
\end{equation} 
with the scalar product defined as 
\begin{equation}
(\mathbf{\Phi}_n, \mathbf{\Phi}_m)=\sum_k \frac{ \Phi_{n,k}  \Phi_{m,k}} {\Phi_{0,k}}=\delta_{n,m},
\label{scal:prod}
\end{equation} 
and 
\begin{equation}
c_m=(\mathbf{p}(0), \mathbf{\Phi}_m).
\label{cn:def}
\end{equation} 
Here the letter $n$ ($m$) enumerates the eigenvector corresponding to a given eigenvalue $\lambda_n$ ($\lambda_m$), while the letter $k$ enumerates the component of  the vector $\bp$.
We remind the reader that the eigenvalues are enumerated in decreasing order $\lambda_0=0>\lambda_1 \ge \lambda_2 \dots \lambda_{N-1}$.

We notice that the eigenvectors  $\mathbf{\Phi}_n$ and eigenvalues $\lambda_n$  of  $\bW$ only depend on the temperature of the bath governing the dynamics, and for the case considered in this paper it is the final temperature $\betafin$.
This implies that in the above expansion \eqref{pi:eigenve} the first eigenvector is 
\begin{equation}
\mathbf{\Phi}_0=\bpeq(\betafin).
\end{equation} 
We also notice that throughout this paper for the initial state  we have chosen $\bp(0)=\bpeq(\betain)$.  Thus the coefficients $c_m$ introduced in \eqref{cn:def} depend on both $\betain$ and $\betafin$.

We now consider the KL divergence for the two possible processes introduced in eq.~\eqref{KL:def1}, 
and obtain
\begin{equation}
D(\bp_i^f(t)|| \bpeq_f)= \sum_k p_k(t) \pq{\ln p_k(t)-\ln \Phi_{0,k}}
\end{equation} 
In order to study the long time behaviour of this  quantity we first consider 
\begin{eqnarray}
&&\log p_k(t)=\log \Phi_{0,k} + \log\p{1+ \sum_{n=1}^{N-1}  c_n \E^{\lambda_n t } \frac{\Phi_{n,k}}{\Phi_{0,k}}}\nonumber \\
&&=\log \Phi_{0,k} + \sum_{l=1}^\infty \frac{ (-1)^{l+1}}{l} \p{ \sum_{n=1}^{N-1}  c_n \E^{\lambda_n t } \frac{\Phi_{n,k}}{\Phi_{0,k}}}^l,
\end{eqnarray}

Now in the long time limit we can approximate the above expression by only taking the term $l=1$ in the  Taylor series, giving
\begin{equation}
\log p_k(t)\simeq \log \Phi_{0,k}  + \sum_{n=1}^{N-1}  c_n \E^{\lambda_n t } \frac{\Phi_{n,k}}{\Phi_{0,k}},
\end{equation} 
 thus we can write 
\begin{eqnarray}
&&D(\bp_i^f(t)|| \bpeq_f)\simeq \sum_k  \p{\sum_{n=1}^{N-1}  c_n  \E^{\lambda_n t }   \Phi_{n,k} \sum_{m=1}^{N-1}  c_m \E^{\lambda_m t } \frac{\Phi_{m,k}}{\Phi_{0,k}} }\nonumber\\
&&=\sum_{n=1}^{N-1}  c_n^2  \E^{2 \lambda_n t }
\end{eqnarray} 
which is clearly dominated by the second eigenvalue $\lambda_1$.
In obtaining the last result we have used the orthonormality of the eigenvectors eq.~(\ref{scal:prod}).

We are also interested in the Fisher information which in terms of the expansion~\eqref{pi:eigenve} can be written as  
\begin{equation}
I(t)=\sum_k \frac{(\partial_t p_k(t))^2}{p_k(t)}=\sum_k \frac{\p{\sum_{n=1}^{N-1} \lambda_n  c_n \E^{\lambda_n t } \Phi_{n,k}}^2}{\Phi_{0,k}+ \sum_{n=1}^{N-1}  c_n \E^{\lambda_n t } \Phi_{n,k}}
\label{I:Fish}
\end{equation} 
and taking the leading terms in the long time limit one obtains
\begin{eqnarray}
&&I(t)\simeq \sum_k  \p{\sum_{n=1}^{N-1}  c_n \lambda_n \E^{\lambda_n t } \frac{\Phi_{n,k}}{\Phi_{0,k}}\,  \sum_{m=1}^{N-1}  c_m \lambda_m \E^{\lambda_m t }\Phi_{m,k}}\nonumber\\
&&=\sum_{n=1}^{N-1}  c_n^2 \lambda_n^2 \E^{2 \lambda_n t }.
\label{long:I}
\end{eqnarray} 
This expression is also dominated by the second eigenvalue $\lambda_1$ of the stochastic matrix, and one finds that in the long time limit the following equality holds $\partial_t^2D(\bp_i^f(t)|| \bpeq_f)=I_i^f(t)/4$.\\

\section{Symmetry of the statistical length}
\subsection{$N=2$}
\label{symm:N2}
For the case of $N=2$ states, the expression for the statistical length can be obtained analytically.
We have 
\begin{equation}
p_k(t)= \Phi_{0,k}+   c_1 \E^{\lambda_1 t } \Phi_{1,k},\qquad i=0,\, 1
\end{equation} 
with $\lambda_1=-w^\up (1+\exp(-\beta \epsilon))$, $\mathbf{\Phi}_1=(-1,1)/\mathcal N_1$, $\mathcal N_1=2 \cosh(\beta \epsilon/2)$.
The Fisher information reads
\begin{eqnarray}
I(t)&=&\sum_k \frac{\p{\partial_t p_k(t)}^2 }{p_k(t)}=\sum_k\frac{ (c_1 \lambda_1 \E^{-\lambda_1 t } \Phi_{1,k})^2}{\Phi_{0,k}+   c_1 \E^{-\lambda_1 t } \Phi_{1,k}}\nonumber \\
&=& \frac{ \lambda_1^2 x_1^2(t)}{(\Phi_{0,0}-x_1(t) ) (\Phi_{0,1} + x_1(t))}, 
\label{It:2}
\end{eqnarray} 
with 
\begin{equation}
x_1(t)=\frac{ c_1 \E^{-\lambda_1 t }} {\mathcal N_1}.
\label{x1:def}
\end{equation} 
Thus we have for the statistic length
\begin{eqnarray}
\mathcal{L}(t)&=&\int_0^t \frac{\lambda_1 |x_1(\tau)| \D \tau}{\sqrt{(\Phi_{0,0}-x_1(t) ) (\Phi_{0,1} + x_1(t))}}\nonumber \\
&=&\pm \int_0^t \frac{\lambda_1 x_1(\tau) \D \tau}{\sqrt{(\Phi_{0,0}-x_1(t) ) (\Phi_{0,1} + x_1(t))}}\nonumber \\
&=&\pm \int^{x_1(0)}_{x_1(t)} \frac{ \D x}{\sqrt{(\Phi_{0,0}-x ) (\Phi_{0,1} + x)}}
\end{eqnarray} 
where the sign in the last expression depends on the sign of $c_1$ in eq.~(\ref{x1:def}).  

From eq.~(\ref{cn:def}) we obtain
\begin{eqnarray}
c_1(-\to +)&=&(\bpeq(\beta_-), \mathbf{\Phi}_1(\beta_+)) = \E^{-\epsilon\beta_-/2}  \frac{\E^{\epsilon(\beta_- -\beta_+)}-1}{1+ \E^{-\epsilon\beta_+}}>0 \nonumber \\
c_1(+\to -)&=&(\bpeq(\beta_+), \mathbf{\Phi}_1(\beta_-)) = \E^{-\epsilon\beta_+/2}  \frac{\E^{-\epsilon(\beta_- -\beta_+)}-1}{1+ \E^{-\epsilon\beta_-}}<0\nonumber 
\end{eqnarray} 
and thus 
\begin{widetext}
\begin{eqnarray}
\mathcal{L}^-_+(t)=2 \pq{\arcsin\sqrt{ \frac{\E^{\epsilon \beta_-} + \E^{\epsilon(\beta_- +\beta_+)}+  \E^{-\lambda^-_1 t  }( \E^{\epsilon\beta_+} - \E^{\epsilon\beta_-} )}{( 1+\E^{\epsilon\beta_+} )(1+ \E^{\epsilon\beta_-} )} }-\arcsin\p{ \p{1+ \E^{-\epsilon \beta_+}}^{-1/2} }},\nonumber \\
\mathcal{L}^+_-(t)=-2 \pq{\arcsin\sqrt{ \frac{\E^{\epsilon \beta_+} + \E^{\epsilon(\beta_+ +\beta_-)}+  \E^{-\lambda^+_1 t  }( \E^{\epsilon\beta_-} - \E^{\epsilon\beta_+} )}{( 1+\E^{\epsilon\beta_+} )(1+ \E^{\epsilon\beta_-} )} }-\arcsin\p{ \p{1+ \E^{-\epsilon \beta_-}}^{-1/2}} }\nonumber,
\end{eqnarray} 
which simplifies as 
\begin{eqnarray}
\mathcal{L}^-_+(t)=2 \pq{\arctan\p{\frac{( 1+\E^{\epsilon\beta_+} )(1+ \E^{\epsilon\beta_-} )}{1+\E^{\epsilon \beta_+}+  \E^{-\lambda^-_1 t  }( \E^{\epsilon\beta_-} - \E^{\epsilon\beta_+} )} -1}^{1/2}-\arctan\p{ \E^{\epsilon \beta_+/2}} },\nonumber \\
\mathcal{L}^+_-(t)=-2 \pq{\arctan\p{\frac{( 1+\E^{\epsilon\beta_+} )(1+ \E^{\epsilon\beta_-} )}{1+\E^{\epsilon \beta_-}+  \E^{-\lambda^+_1 t  }( \E^{\epsilon\beta_+} - \E^{\epsilon\beta_-} )} -1}^{1/2}-\arctan\p{ \E^{\epsilon \beta_-/2}} },\nonumber 
\end{eqnarray}

thus exhibiting the symmetry $\mathcal{L}^-_+(\infty)=\mathcal{L}^+_-(\infty)$.
\end{widetext}

We also notice that from eq.~(\ref{It:2}), one obtains the following inequality for the initial values of the Fisher information
\begin{equation}
I_-^+(0)-I_+^-(0)= \E^{-2 (\beta_++\beta_-)\epsilon} \p{\E^{\beta_-\epsilon} -\E^{\beta_+\epsilon}}^3\ge 0
\end{equation} 
for $\beta_-\ge \beta_+$.

\subsection{$N>2$}
\label{symm:NN}
For a system with an arbitrary number of states in general there is not the symmetry  $\mathcal{L}^-_+(\infty)=\mathcal{L}^+_-(\infty)$.
This is  corroborated by the numerical findings of the main text.
In the following we show that the symmetry of the statistical length is restored in the limit of small $\epsilon$.
In this limit we notice that we can write
\begin{equation}
\frac{p_k(0)}{\Phi_{0,i}}=\pq{1 + \p{\frac{N-1}2 -i} \epsilon (\betain -\betafin)}+ O(\epsilon^2).
\end{equation} 

Introducing the expansions in $\epsilon$:  $\lambda_n=\lambda_n^{(0)} + \epsilon \lambda_n^{(1)}+ \dots$ and $\Phi_{n,i}= \Phi_{n,i}^{(0)}+ \epsilon \Phi_{n,i}^{(1)}+ \dots$, from  eqs.~\eqref{scal:prod}--\eqref{cn:def} we obtain for $c_n$ 
\begin{eqnarray}
&&c_n=\sum_k \Phi_{n,k}+ \epsilon (\betain -\betafin) \sum_k \p{\frac{N-1}2 -k}\Phi_{n,k}+ \dots\nonumber \\
&&= \epsilon (\betain -\betafin) \sum_k \p{\frac{N-1}2 -k}\Phi_{n,k}^{(0)}+ O(\epsilon^2),
\label{cn:exp}
\end{eqnarray} 
where we have used the orthonormality condition eq.~(\ref{scal:prod}).
Using this last result to analyze eq.~(\ref{I:Fish}), one realizes that the leading term in $\epsilon$ in the Fisher information arises from the expansion of $c_n$ (\ref{cn:exp}), and obtains the result
\begin{equation}
I=\epsilon^2 (\betain-\betafin)^2 \p{\sum_n \tilde c_n \lambda_n^{(0)} \E^{-\lambda_n^{(0)} t }}^2  + O(\epsilon^3),
\end{equation} 
where 
\begin{equation}
\tilde c_n=\sum_k k \cdot \Phi_{n,k}^{(0)}.
\end{equation} 
%\Phi_{0,i}+ \sum_{n=1}^N  c_n \E^{-\lambda_n t } \Phi_{n,i},
where neither $\tilde c_n$ nor $\lambda_n^{(0)}$ depend on the temperature.

Thus the statistical length for $t\to\infty$ reads
\begin{eqnarray}
&&\mathcal L(\infty)=\int_0^\infty \D \tau \sqrt{I(\tau)}\nonumber\\
&&= \epsilon |\betain-\betafin|  \int_0^\infty \D \tau \left| \sum_n \tilde c_n \lambda_n^{(0)} \E^{-\lambda_n^{(0)} \tau} \right| \nonumber \\
&&\quad + O(\epsilon^2).
\label{eq:L:small:eps}
\end{eqnarray} 
We conclude that to the lowest order in $\epsilon$ the statistical length is symmetric. This is in agreement with the finding of \cite{Ibanez2024} that for the classical harmonic oscillator with Brownian dynamics  $\mathcal{L}^-_+(\infty)=\mathcal{L}^+_-(\infty)$.  In the main text we provide numerical evidence  that for finite $\epsilon$ this symmetry disappears.
\section{Characteristic polynomial and eigenvalues of the stochastic matrix (\ref{W:eq})}
\label{eigenv:app}
The characteristic polynomial of the $N\times N$ stochastic matrix (\ref{W:eq})
can be obtained through the Chebyshev recursion relations for tridiagonal matrices, resulting in the expression 

\begin{eqnarray}
&&P_N(\lambda)=\frac{ (-1)^N \lambda} {2^N\sqrt{ (a + \lambda)^2-4 b c }}\nonumber \\
&& \times \pq{(a + \lambda + \sqrt{(a + \lambda)^2-4 b c })^
   N - (a + \lambda - \sqrt{ (a + \lambda)^2-4 b c })^N}\nonumber,
\label{P:lambda}
\end{eqnarray} 
with $b=w^\up$, $c=w^\down$, and $a=b+c$.
One immediately sees that $\lambda=0$ is an eigenvalue. 
A  tedious but straightforward calculation leads to the remaining $N-1$ eigenvalues as given by eq.~\eqref{lambda:N}.

\bibliography{bibliography}

%\begin{thebibliography}{}
%\bibitem{Ibanez23} Miguel Ib\'a\~nez {\it et al.} arXiv:2302.09061
%%\bibitem{Hans:notes} Hans Fogedby, private communication.
%\end{thebibliography}

\end{document}